\documentclass[nojss]{jss}

\usepackage{thumbpdf,lmodern}

\usepackage{framed}

\newcommand{\class}[1]{`\code{#1}'}
\newcommand{\fct}[1]{\code{#1()}}
\newcommand{\TT}{\mathbf{T}}
\newcommand{\lo}{\mathrm{low}}
\newcommand{\up}{\mathrm{upp}}
\newcommand{\Bbeta}{{\boldsymbol{\beta}}}
\newcommand{\Bgamma}{{\boldsymbol{\gamma}}}
\providecommand{\1}{\mathbf{1}}
\newtheorem{definition}{Definition}[section]

\usepackage{amsmath,amsfonts,amssymb}

\makeatletter
\setkeys{Gin}{width=\Gin@nat@width}
\makeatother

\author{Mari Myllym{\"a}ki\\Natural Resources Institute Finland (Luke)
   \And Tom\'a\v s Mrkvi\v cka\\University of South Bohemia}
\Plainauthor{Mari Myllym{\"a}ki, Tom\'a\v s Mrkvi\v cka}

\title{\pkg{GET}: Global Envelopes in \proglang{R}}
\Plaintitle{GET: Global Envelopes in R}
\Shorttitle{\pkg{GET}: Global Envelopes in \proglang{R}}

\Abstract{
This work describes the \proglang{R} package \pkg{GET} that implements global
envelopes for a general set of $d$-dimensional vectors $T$ in various applications.
A $100(1-\alpha)$\% global envelope is a band bounded by two vectors such that
the probability that $T$ falls outside this envelope in any of the $d$ points is
equal to $\alpha$.
The term 'global' means that this probability is controlled simultaneously
for all the $d$ elements of the vectors.
The global envelopes can be employed for central regions of functional or multivariate data,
for graphical Monte Carlo and permutation tests where the test statistic is multivariate or functional,
and for global confidence and prediction bands.
Intrinsic graphical interpretation property is introduced for global envelopes.
The global envelopes included in the \pkg{GET} package have this property,
which particularly helps to interpret test results, by providing a graphical interpretation
that shows the reasons of rejection of the tested hypothesis.
Examples of different uses of global envelopes and their implementation in the \pkg{GET} package are presented,
including global envelopes for single and several one- or two-dimensional functions,
Monte Carlo goodness-of-fit tests for simple and composite hypotheses,
comparison of distributions,
functional analysis of variance,
and functional linear model,
and confidence bands in polynomial regression.
}

\Keywords{functional linear model, central region, goodness-of-fit, graphical normality test, Monte Carlo test, multiple testing, permutation test, \proglang{R}, spatial point pattern}
\Plainkeywords{functional linear model, central region, goodness-of-fit, graphical normality test, Monte Carlo test, multiple testing, permutation test, R, spatial point pattern}

\Address{
  Mari Myllym{\"a}ki\\
  Natural Resources Institute Finland (Luke)\\
  Latokartanonkaari 9\\
  FI-00790 Helsinki, Finland\\
  E-mail: \email{mari.myllymaki@luke.fi}\\
  URL: \url{https://www.luke.fi/en/experts/mari-myllymaki/} \\
  \emph{and}\\
  Tom\'a\v s Mrkvi\v cka \\
  Dpt. of Applied Mathematics and Informatics \\
  Faculty of Economics \\
  University of South Bohemia, \\
  Studentsk{\'a} 13 \\
  37005 \v{C}esk\'e Bud\v{e}jovice, Czech Republic \\
  E-mail: \email{mrkvicka.toma@gmail.com} \\
  URL: \url{http://home.ef.jcu.cz/~mrkvicka/}
}

\begin{document}


\section{Introduction}\label{sec:intro}

Global envelopes are useful for formal testing of various hypotheses using functional or multivariate statistics when interpretation of the test results is of key interest, for determining central regions of functional or multivariate data, and also for determining confidence or prediction bands \citep[e.g.,][]{MyllymakiEtal2017, MrkvickaEtal2017, NarisettyNair2016}. Global envelopes have shown their usefulness already in many areas, e.g., spatial statistics, 
functional data analysis and 
image analysis, 
with applications to
agriculture, 
architecture and art, 
astronomy and astrophysics, 
ecology, 
evolution, 
economics, 
eye movement research, 
fisheries, 
forestry, 
geography, 
material science, and 
medicine, health and neurosciences \citep[e.g.,][]{Stoyan2016, Murrell2018, HabelEtal2017, ChaibanEtal2019, PollingtonEtal2020}. 
To make these methods easily accessible, the R \citep{R2020} package \pkg{GET} has been developed that is available from the Comprehensive R Archive Network (CRAN) at \url{https://cran.r-project.org/package=GET}. A development version of the package is available via the repository \url{https://github.com/myllym/GET}. The package provides an implementation of global envelopes in various settings.

Because there are many other methods that can be used for the same purposes as global envelopes, we first give a motivating example of functional analysis of covariance (ANCOVA) to demonstrate the main features and advantages of global envelope tests (Section~\ref{sec:motivating_example}). Other possible usages are also discussed in more detail.
The second part of this introductory section describes the competing and complementary methods and software for these same usages.
Thereafter, in Section~\ref{sec:implementation}, we give an overview of global envelopes including the formal definition of their graphical interpretation, summary of the types of global envelopes and their implementation in \pkg{GET}.
In Section~\ref{sec:illustrations}, the usage of global envelopes is shown for several examples of applications illustrating main features of the \pkg{GET} package, namely 1) the computation of central regions and functional boxplots for a set of functions or jointly for several sets of functions (Section~\ref{sec:cr}); 2) the Monte Carlo goodness-of-fit test for simple hypotheses with application to spatial statistics (Section~\ref{sec:GOT}); 3) the Monte Carlo goodness-of-fit test for composite hypotheses with application to graphical normality testing (Section~\ref{sec:GOTcomposite}); 4) the graphical $n$-sample test of correspondence of distribution functions, $n\geq 2$ (Section~\ref{sec:CDF}); 5) the graphical functional one-way analysis of variance (ANOVA) (Section~\ref{sec:fANOVA}); 6) the functional general linear model (GLM) for images (Section~\ref{sec:fGLM}); and 7) the computation of the confidence band in polynomial regression (Section~\ref{sec:confband}).
The final section, Section~\ref{sec:summary}, is left for discussion.

\subsection{A motivating example and possible usages}\label{sec:motivating_example}

\begin{figure}[ht!]
\centering
\includegraphics{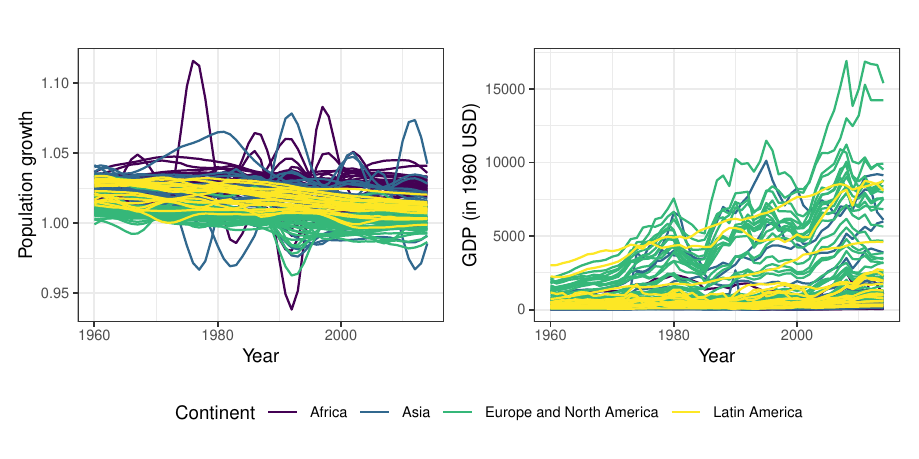}
  \caption{Population growth and GDP (in 1960 USD) curves for 1960--2014 for 112 countries from the four continents.}
  \label{fig:popgrowth}
\end{figure}

Let us study the differences between population growth accumulated in different continents over the period of 55 years. We assume that GDP (gross domestic product) can be the main driver of population growth in different countries, therefore we add GDP into the model as a nuisance factor.
Figure~\ref{fig:popgrowth} (left) shows population growth, defined here as the population at the end of the year divided by the population at the beginning of the year, in years from 1960 to 2014 in Africa, Asia, Europe and North America, and Latin America, total in 112 countries with more than one million inhabitants in 1950 \citep{NagyEtal2017,DaiEtal2020}. Figure~\ref{fig:popgrowth} (right) shows the GDP of every country in the study discounted to the 1960 USD, according to USD inflation. The data was obtained from the World Bank. The missing values of the GDP of a country were extrapolated using the closest known ratio of the GDP of the country and the median GDP in that year, and interpolated using linear interpolation of the two closest ratios.

Since the global envelopes are a nonparametric tool, we can use as a test statistic any multivariate or functional test statistic without the worry of breaking assumptions of normality or homogeneity of the distribution of the test statistic in different years. Also, since we can use any test statistic, we can combine several functional test statistic into one (for more details about combining test statistics see Appendix~\ref{app:combinedGE}). Here, we combine four functional test statistics, each representing a coefficient vector attached to a continent in our ANCOVA model, i.e., our test statistic is
\begin{eqnarray}\label{TT1}
\nonumber &\Bbeta^{\text{Cont}} = (\underbrace{\beta_{1, 1960}^{\text{Cont}}, \ldots , \beta_{1, 2014}^{\text{Cont}}}_{1:\, \text{Africa}}, 
     \underbrace{\beta_{2, 1960}^{\text{Cont}}, \ldots , \beta_{2, 2014}^{\text{Cont}}}_{2:\, \text{Asia}},\\ & 
     \underbrace{\beta_{3, 1960}^{\text{Cont}}, \ldots , \beta_{3, 2014}^{\text{Cont}}}_{3:\, \text{Europe and North America}}, 
     \underbrace{\beta_{4, 1960}^{\text{Cont}}, \ldots , \beta_{4, 2014}^{\text{Cont}}}_{4:\, \text{Latin America}}),
\end{eqnarray}
where $\beta_{1,i}^{\text{Cont}}, \ldots \beta_{4,i}^{\text{Cont}}$ are the parameters of the univariate ANCOVA model for year $i$, computed with condition $\sum_{j=1}^4 \beta_{j,i}^{\text{Cont}} = 0$. This model parametrization enables that the parameter  $\beta_{j,i}^{\text{Cont}}$ is the difference between the $j$-th continent's growth rate and the overall mean.

\begin{figure}[ht!]
\centering
\includegraphics{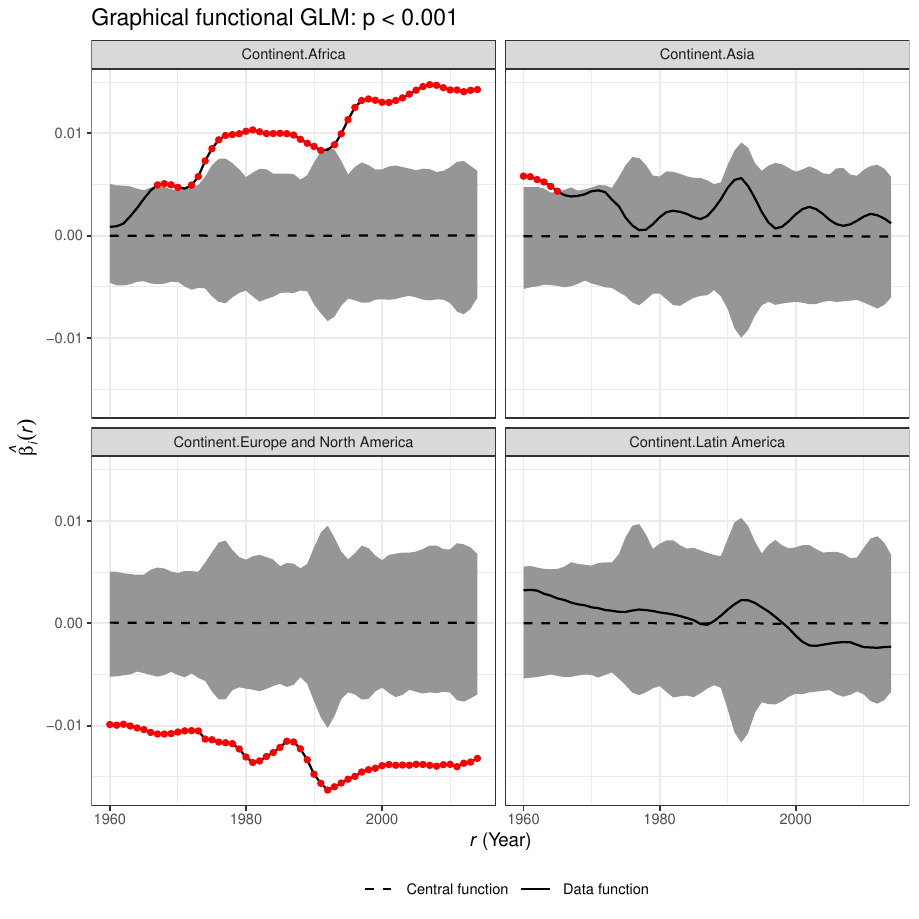}
  \caption{Test for the main effect of the continent, given the GDP (in 1960 USD) as a nuisance factor. The red color is used to highlight the (significant) years where the data coefficient $\hat{\beta}_{j,i}^{\text{Cont}}$, $i=1960, \ldots, 2014$, $j=1,2,3,4$, goes outside the 95\% global envelope computed under the null model of no continent effect. Here $j$ refers to the continent given in the titles.}
  \label{fig:popgrowth_flm}
\end{figure}

This test statistic allows for a direct comparison of each continent's growth rate with the world's mean growth rate. Figure~\ref{fig:popgrowth_flm} shows the test statistic (Equation~\ref{TT1}) computed for the data together with 95\% global envelope computed under the null model of no continent effect but with nuisance effect of GDP. This output shows, first, the formal global $p$~value, which indicates that the effect of the continent is significant after applying an adjustment for multiple comparisons for the 55 years, and four continents. Second, it shows the reasons of rejections, i.e., which years have led to the significant test result. Due to the intrinsic graphical interpretation (IGI, see Definition~\ref{IGI}) of the global envelope, we have a one-to-one correspondence between the formal global test and its graphical interpretation, i.e., the data statistic lies outside the envelope for some point (here years and continents) if and only if the global test is significant. This allows for direct identification of the significant years. Third, the output shows the direction and amount of deviation from the null hypothesis. For example, the growth rate in Europe and North America is significantly lower than the world's average in all years, whereas Asia's growth rate is slightly significantly higher than the world's average only up to the year 1966. Fourth, due to the free choice of the test statistic, the test output shows here the comparison individually for every continent (level of the categorical predictor). The \pkg{GET} package also allows for the choice of comparison between all pairs of continents in the style of the post hoc test. Note that usually, the two tests are applied - global AN(C)OVA and multiple comparison post hoc test.
Note also that we can calculate results also for continuous GDP effect and interaction effect between the GDP and all continents (all levels of the categorical effect) in the same way \citep[see][Section~6]{MrkvickaMyllymaki2023}.

Technically, the global envelope of Figure~\ref{fig:popgrowth_flm} was computed from 5000 permutations under the null model using the standard \citet{FreedmanLane1983} permutation procedure. The IGI measure used was the area measure (see Appendix~\ref{app:GEdef} for details).

The above example illustrates the global envelope test, a possible usage of global envelopes.
The usage of global envelopes is however more versatile. They can be used for producing
\begin{enumerate}
\item[(i)] a central region: a central region is constructed for a set of vectors or functions in order to find central or outlying vectors or functions (e.g., outlier detection, functional boxplot);
\item[(ii)] a global envelope test: a Monte Carlo goodness-of-fit test or a permution test where the test statistic is multivariate or a function of any dimension (e.g., goodness-of-fit test for point patterns, random sets, or for a family of distributions, functional ANOVA, functional GLM, $n$-sample test of correspondence of distribution functions);
\item[(iii)] global confidence or prediction bands: a confidence or prediction band is produced from a set of vectors or functions obtained by bootstrap or sampling from Bayesian posterior distribution (e.g., confidence band in polynomial regression, Bayesian posterior prediction).
\end{enumerate}
In fact, the global envelope test uses the 95\% central region (case (i) usage) computed under the null hypothesis for determining the test results. The global confidence bands (case (iii) usage) also uses the 95\% central region computed under the full model. Thus in core of all applications lies the central region. Central regions can be constructed based on any functional ordering, but we concentrate only in those satisfying the IGI, the one-to-one correspondence of a functional ordering and its graphical interpretation. In fact, when central region is not satisfying IGI and we see a function which is not contained completely inside the central region, we still do not know if such a function can be regarded as extremal. On the contrary, when we use IGI ordering, we know that such a function is among the most, say 5\%, extremal functions.

In this article, we primarily focus on global envelopes for testing the global hypothesis under the control of family-wise error rate (FWER). This is to determine if at least one component of the test statistic can be considered significant. Alternatively, one can be interested in the local test, which aims to differentiate between the domain where the test should be rejected and where it should not be rejected. For this purpose, the false discovery rate (FDR) is typically more suitable for controlling the global level of the multiple test. We have recently developed the FDR envelope \citep{MrkvickaMyllymaki2023}, which accomplishes this task with an envelope which has the IGI under the FDR control: the test is rejected if and only if the data statistic falls outside the envelope. Examples of the FDR envelopes are provided in the vignette which can be accessed by typing \code{library("GET")} and \code{vignette("FDRenvelopes")} in \proglang{R}.

\subsection{Competing and complementary methods and software}\label{sec:intro_software}

Below are listed the other \proglang{R} packages (or code) that we know to provide functions for some global envelopes or central regions.
Further, as already mentioned, the problems (i)-(iii) can be solved by other methods as well, not just by global envelopes. The relation of these methods to the global envelope methods are also discussed below.

{\bf Global envelopes and central regions:}
\begin{itemize}
\item The \proglang{R} package \pkg{fda} \citep{fda2022} provides the function \fct{fbplot} for the computation of the central region and functional boxplot according to two different orderings than those described here, namely the band depth and modified band depth (MBD) \citep{Lopez-PintadoRomo2009, SunEtal2012}, but these depths do not allow for IGI.
\item The \proglang{R} package \pkg{depthTools} \citep{depthTools2013, TorrenteEtal2013} similarly allows for central regions based on MBD (no IGI).
\item The \proglang{R} package \pkg{spatstat} \citep{spatstat2015} provides the function \fct{envelope} for the simulation of envelopes based on a given summary function of a spatial point pattern. By default, \fct{envelope} provides a pointwise envelope, but the option \code{global = TRUE} allows one to compute the global envelope of \citet{Ripley1981}, which corresponds to the \code{'unscaled'} envelope in \pkg{GET} (see Table~\ref{tab:GEoverview}). It has been shown that this \emph{unscaled} global envelope test has generally lower power than the other methods of Table~\ref{tab:GEoverview} \citep{MyllymakiEtal2015, MyllymakiEtal2017}. The corresponding adjusted unscaled global envelope \citep{DaoGenton2014, BaddeleyEtal2017} for composite hypotheses is also provided in \pkg{spatstat} (the function \fct{dg.envelope}).
\item \citet{AldorNoimanEtal2013} presented a global envelope for a Q-Q plot (and provided a link to an \proglang{R} script). The shape of the envelope is derived theoretically, but the size of the envelope has to be computed from simulations. The methods of the \pkg{GET} package can be used for this purpose as well, both for simple and composite hypotheses, but the theoretical achievement of \citet{AldorNoimanEtal2013} for simple hypotheses has apparent advantages.
\item The \proglang{R} package \pkg{boot} \citep{boot2017,DavisonHinkley1997} provides the function \fct{envelope} for the computation of a global envelope from bootstrapped functions. This envelope has the same shape as the global rank envelope (\code{'rank'} in Table~\ref{tab:GEoverview}), but the appropriate envelope ($l$ of Equation~\ref{eq:kth_envelopes}) is chosen in \pkg{boot} experimentally \citep{DavisonHinkley1997}. Since the differences in the nominal levels of the subsequent ($l$-)envelopes from which the choice is made can be large, the predetermined level is reached only approximately.
\item The package \pkg{dbmss} \citep{MarconEtal2015} provides similar global envelopes as the \pkg{boot} package \citep{DurantonOverman2005} but for the global confidence envelopes of spatial summaries.
\item There are other \proglang{R} packages with the ability to compute simultaneous confidence bands for various models, e.g., \pkg{excursions} \citep{BolinLindgren2015,BolinLindgren2017,BolinLindgren2018} 
for Gaussian processes, \pkg{AdaptFitOS} \citep{WiesenfarthEtal2012} for semiparametric regression models and \pkg{SCBmeanfd} \citep{SCBmeanfd2016} for nonparametric regression models with functional data using a functional asymptotic normality result.
Instead, the global envelopes of the \pkg{GET} package (see Table~\ref{tab:GEoverview}) are constructed non-parametrically from a set of vectors.
Apparently, the package \pkg{excursions} allows also for a non-parametric version with the same shape as the global rank envelope (\code{'rank'} in Table~\ref{tab:GEoverview}), similarly as \pkg{boot}.
\end{itemize}

{\bf Multiple testing:}
The global envelope tests can be seen as a general solution to the
multiple testing problem in Monte Carlo tests \citep{MrkvickaEtal2017}.
There are several other methods and \proglang{R} packages to the multiple testing problem controlling FWER.
The few packages mentioned below have a link to the methods of \pkg{GET}:
\begin{itemize}
\item The \proglang{R} packages \pkg{coin} \citep{coin2008, HothornEtal2006} and \pkg{multtest} \citep{PollardEtal2005} enable one to compute the $p$~value adjusted for multiple testing in a multiple permutation test based on the minimum $p$~value computed from all individual tests. The null distribution of the minimum $p$~values or the maximum of a test statistic is obtained from permutations. The minimum $p$~value method corresponds to the conservative rank test based on the $p_+$~value (see global rank envelope in Appendix~\ref{app:GEdef}).
\item General multiple test procedures are also provided by the package \pkg{sgof} \citep{sgof2016} for goodness-of-fit testing and by the package \pkg{stats} \citep{R2020} for adjusting the $p$~values for multiple comparisons by Bonferroni type methods (the function \fct{p.adjust}).
\end{itemize}

{\bf Functional GLM:}
The global envelope tests can also be used for functional GLM using a permutation strategy to generate samples under the null hypothesis. There are several other methods and software to the functional GLM problem: 
\begin{itemize}
\item The PALM software \citep{WinklerEtal2014} allows for the computation of various functional GLM designs using permutation tests. The multiple testing problem is solved by an unscaled envelope constructed for the test statistic (e.g., $F$~statistic).
\item The \proglang{R} packages \pkg{fda.usc} \citep{fdausc2012} and \pkg{fdANOVA} \citep{fdANOVA2017} allow for the computation of functional ANOVA designs by several methods together with the computation of factor's significances. Similarly the package \pkg{fda} \citep{fda2022} allows for computations in functional regression designs. However, to the best of our knowledge, these do not provide the IGI of tests of a factor significance.
\end{itemize}


\section[Overview of global envelopes in GET]{Overview of global envelopes in \pkg{GET}}\label{sec:implementation}

\subsection{Global envelopes and intrinsic graphical interpretation (IGI)}

Global envelopes are constructed for general multivariate statistics, so in the case when the data are purely functional, they first have to be discretized. The discretization of the functions can be arbitrary, as long as it is the same for each function.
Therefore, let $\TT_1, \TT_2, \ldots, \TT_{s}$ be $d$-dimensional vectors, 
$\TT_i = (T_{i1}, T_{i2}, \ldots, T_{id})$ for $i=1,\ldots, s$.
In the case (i) (see Section~\ref{sec:motivating_example}), the vectors $\TT_i$ are observed functions assumed to follow the same distribution and the aim is to find the least extreme $100(1-\alpha)$\% of the $s$ vectors to construct the central region, with $\alpha\in(0,1)$.
In the case (ii), the vector $\TT_1$ is an observed function and $\TT_2,\ldots,\TT_s$ are vectors simulated under the tested null hypothesis. 
The key idea of a global envelope test is the same as a general Monte Carlo or permutation test:
the rank of the data vector $\TT_1$ among all the vectors $\TT_1,\ldots,\TT_s$ determines the result and the $p$~value of the test \citep{Barnard1963}. And, in the case (iii), $\TT_i$ are generated under the given bootstrap or Bayesian scheme.
As mentioned already, in the core of all cases (i)-(iii) lies the central region and, for the purpose of ordering the $d$-dimensional vectors $\TT_i$ (or functions) from the most extreme to the least extreme and finding the central region, many different measures exist. The \pkg{GET} package focuses on such measures for which it is possible to construct the global envelope with a practically interesting graphical interpretation, which we call IGI.

In general, an {\em envelope} is considered to be a band bounded by two vectors.
A $100(1-\alpha)$\% {\em global envelope} is a set $(\TT_{\text{low}}^{(\alpha)}, \TT_{\text{upp}}^{(\alpha)})$ of envelope vectors with
$\TT_{\text{low}}^{(\alpha)}=(T_{\text{low}\,1}^{(\alpha)}, \ldots, T_{\text{low}\,d}^{(\alpha)})$ and 
$\TT_{\text{upp}}^{(\alpha)}=(T_{\text{upp}\,1}^{(\alpha)}, \ldots, T_{\text{upp}\,d}^{(\alpha)})$ 
such that the probability that $\mathbf{T}_i$ falls outside this envelope in any of the $d$ points is equal to $\alpha$, for $\alpha\in (0,1)$, i.e.,
\begin{equation}\label{eq:global_envelope}
P(T_{ik} \notin [T_{\lo\,k}^{(\alpha)}, T_{\up\,k}^{(\alpha)}] \text{ for any } k\in\{1,\ldots,d\}) = \alpha.
\end{equation}
In all cases (i)-(iii), {\it global} means that the envelope is given with the prescribed coverage $100(1-\alpha)$\% simultaneously for all the elements of the multivariate or functional statistic, but the probability depends on the situation (i)-(iii).
In case (i), the probability is taken under the distribution of the vectors $\TT_i$.
In case (ii), the probability is taken under the null hypothesis $H_0$,
and, in case (iii), the probability is taken under the distribution of the random vector $\TT_i$ generated under the given bootstrap or Bayesian scheme.
It should be noted that in a {\em pointwise} (or local) envelope the probability to fall out of the envelope is controlled instead individually for every element of the vector $\TT_i$.

We define the IGI property for global envelopes as follows.
\begin{definition}\label{IGI} 
Assume that a general ordering $\prec$ of the vectors $\TT_i, i=1, \ldots , s$, is induced by a univariate measure $M_i$. That is,
$M_i \geq M_j$ iff  $\TT_i \prec \TT_j$, which means that $\TT_i$ is less extreme or as extreme as $\TT_j$. (The smaller the measure $M_i$, the more extreme the $\TT_i$ is.) 
The $100(1-\alpha)$\% global envelope $[\TT_{\lo\,k}^{(\alpha)}, \TT_{\up\,k}^{(\alpha)}]$ has intrinsic graphical interpretation (IGI) with respect to the ordering $\prec$ if
\begin{enumerate}
  \item $M_{(\alpha)} \in \mathbb{R}$ is the largest of the $M_i$ such that the number of those $i$ for which $M_i< M_{(\alpha)}$
is less or equal to $\alpha s$;
 \item $T_{ik} < T_{\lo\,k}^{(\alpha)}$ or $T_{ik} > T_{\up\,k}^{(\alpha)}$ for some $k = 1, \ldots , d$
iff $M_i<M_{(\alpha)}$ for every $i=1, \ldots , s$;
 \item $T_{\lo\,k}^{(\alpha)} \leq T_{ij} \leq T_{\up\,k}^{(\alpha)}$ for all $k = 1, \ldots , d$ iff $M_i\geq M_{(\alpha)}$ for every $i=1, \ldots , s$.
\end{enumerate}
\end{definition}
The points 2.\ and 3.\ are equivalent, but we specify them both in order to stress both graphical properties of IGI. The IGI property means that the vector $\TT_i$ is outside the $100(1-\alpha)\%$ global envelope in any of its components if and only if the vector is considered to be extreme by the measure $M$ at the level $\alpha$, and the vector $\TT_i$ is completely inside the global envelope if and only if the vector is not extreme at the level $\alpha$. Thus, the global envelope with IGI provides a solution to the tasks (i)-(iii) in a graphical manner.

In particular in case (ii), the data vector $\TT_1$ is compared with a global envelope in order to decide if the data vector is extreme ($H_0$ is rejected) or not extreme ($H_0$ is not rejected), and to find reasons for the possible rejection of $H_0$ through inspecting for which $k=1,\ldots,d$ the data vector $\TT_1$ is outside the global envelope.
For this testing task, in addition to a global envelope, a Monte Carlo $p$~value is computed according to the measure $M_i$: $p = \sum_{i=1}^{s} \1 (M_i \leq M_1) \big/ s$ \citep[see, e.g.,][]{MyllymakiEtal2017, MrkvickaEtal2017, MrkvickaEtal2020, MrkvickaEtal2022, MrkvickaEtal2021a}.
In order to obtain an exact Monte Carlo test, i.e., a test that achieves the prescribed FWER, the exchangeability of the test vectors $\TT_i$ is required. All the examples in Section~\ref{sec:illustrations} satisfy exchangeability, except the functional GLM where the permutation of the residuals from the null model \citep{FreedmanLane1983} is used. This permutation scheme is commonly used in univariate permutation GLMs and widely accepted as the best available solution \citep{AndersonRobinson2001, AndersonTerBraak2003, WinklerEtal2014}.

\subsection{Types of global envelopes}

The \pkg{GET} package implements seven global envelopes defined in earlier works as specified in Table~\ref{tab:GEoverview}, together with their short descriptions and specifications in the \pkg{GET} functions. Definitions of these envelopes are given in Appendix~\ref{app:GEdef}.
The first four envelopes in Table~\ref{tab:GEoverview}
(\code{'rank'}, \code{'erl'}, \code{'cont'}, \code{'area'}) 
are completely non-parametric envelopes and are called global rank envelopes,
because they are all based on pointwise ranks of the vectors $\TT_i$. The extreme rank length, continuous and area envelopes are refinements to the rank envelope.
Note here that if a global FWER control is provided within other packages, it is only provided for not refined \code{'rank'} envelope or \code{'unscaled'} envelope.
(see details in Appendix~\ref{app:GEdef}). The \code{'st'} and \code{'qdir'} envelopes parameterize the marginal distributions of $T_{i k}$, $i=1, \ldots, s$ by one or two parameters, respectively, for all $k = 1,\ldots, d$. Thus they can be regarded as approximations of the first four envelopes.

In a typical application one needs to choose one of the measures with IGI (Table~\ref{tab:GEoverview}).
In general, the first five types of Table~\ref{tab:GEoverview} instead of the last two, \code{'st'} and \code{'unscaled'}, can be recommended based on previous studies \citep{MyllymakiEtal2015, MyllymakiEtal2017}.
Regarding the choice between the first five types, when one can afford a large number of simulations in cases (ii)-(iii) of Section~\ref{sec:motivating_example}, one can well use type \code{'erl'} that is based only on the ranks, thus also suiting particularly well for combined tests where the test statistic is a combination of several functional test statistics (see Appendix~\ref{app:combinedGE} and examples in Sections~\ref{sec:motivating_example} and \ref{sec:cr}). 
On the other hand, any other choice is also fine, because the \code{'rank'}, \code{'erl'}, \code{'cont'} and \code{'area'} measures lead to an equivalent outcome for a large number of simulations or permutations. However, the definition of {\em large} depends on the situation.
A simulation study presented in \citet{MyllymakiMrkvicka2020b} supplements this article, giving guidance on the required number of simulations under different scenarios.

Another situation arises in case (i) with a low number of vectors or functions, or in cases (ii)-(iii) where the simulations or permutations are too time consuming to have a large number of them. Then the choice of the measure plays a role. Based on our experience supported by the simulation study presented in \citet{MyllymakiMrkvicka2020b}, the \code{'erl'} and \code{'area'} measures are typically good choices for detecting integral type of extremeness where the vector $\TT_i$ is extreme in the set of vectors for a large range of its components. On the other hand, the \code{'cont'} and \code{'qdir'} measures are most sensitive to the maximum type of extremeness, i.e., the case where $\TT_i$ is extreme only for a few of its components, but also the \code{'area'} measure performs well. Thus, if no particular type of extremeness is expected a priori, the \code{'area'} measure is often a good compromise, since it is sensitive to the amount of outlyingness (similarly as \code{'erl'}) and to the value of outlyingness (similarly as \code{'cont'} and \code{'qdir'}). Illustration of the different measures can be found in \citet[][Section~2.4]{MrkvickaEtal2022}.

To conclude, for testing, if just possible, we recommend to use a large number of simulations and one of the first five measures of Table~\ref{tab:GEoverview}. If large number of simulations is not possible, then still Barnard's Monte Carlo test is valid. For testing and ordering functional data, use then the measure which is sensitive to the type of extremeness that you regard as important. If the type of extremeness is unknown, then the \code{'area'} measure can be preferred.

\begin{table}[t!]
\centering
\begin{tabular}{lp{4.0cm}p{9.0cm}}
\hline
Type           & Introduced in   & Description \\ \hline
\code{'rank'}         & \citet{MyllymakiEtal2017}  & Global rank envelope  corresponding to the extreme rank measure (with ties); unique ordering (or $p$~value) provided additionally as specified in the argument \code{ties}, e.g., \code{'erl'} for extreme rank lengths \\
\code{'erl'}          & \citet{MyllymakiEtal2017, NarisettyNair2016, MrkvickaEtal2020}    & Global rank envelope corresponding to extreme rank length (ERL) measure \\ 
\code{'cont'}         & \citet{Hahn2015, MrkvickaEtal2022} & Global rank envelope corresponding to the continuous rank measure \\
\code{'area'}         & \citet{MrkvickaEtal2022}    &  Global rank envelope corresponding to the area measure \\
\code{'qdir'}         & \citet{MyllymakiEtal2017,MyllymakiEtal2015}       & Directional quantile envelope test corresponding to the directional quantile maximum absolute deviation (MAD) measure \\
\code{'st'}           & \citet{MyllymakiEtal2017, MyllymakiEtal2015}      & Studentized envelope test corresponding to the studentized MAD measure \\
\code{'unscaled'}     & \citet{Ripley1981}       & Unscaled envelope test corresponding to the classical, unscaled MAD measure. The envelope has a constant width. \\ \hline
\end{tabular}
\caption{\label{tab:GEoverview} Overview of different types of global envelopes in the \pkg{GET} package. The types \code{'erl'}, \code{'cont'} and \code{'area'} refine the type \code{'rank'} by breaking the ties in the extreme ranks, for details see Appendix~\ref{app:GEdef}. 
}
\end{table}

As it was mentioned earlier, other measures or depths without IGI can be used for the ordering the data. We believe that there exists no specific depth/measure which would perform universally the best and that for every depth/measure it is possible to construct a situation for which the chosen depth/measure will be the most powerful. However, for an application at hand, it is typically not possible to know the most powerful depth/measure; one has to make the choice of the depth/measure in advance. Therefore, we believe that it is often practical to choose the IGI measure, which brings an easier interpretation. In a particular study of the power of goodness of fit tests in spatial statistics our IGI measures were much more powerful than the modified band depth and modified half region depth \citep{MyllymakiEtal2017}, but no universal conclusions can be drawn from a single study.

\subsection[Implementation and key functions in GET]{Implementation and key functions in \pkg{GET}}

As described above, the construction of a global envelope is based on a measure $M$. 
The calculation of different measures in the \pkg{GET} package is provided by the function \fct{forder} (functional ordering).
Most often, the user however calls either \fct{central\_region} for constructing central regions with IGI or \fct{global\_envelope\_test} for performing global envelope tests (equipped with $p$~values as well). Both functions utilize \fct{forder} for the calculation of the measures $M$.
The most important arguments of these functions are
\begin{Code}
central_region(curve_sets, type = 'erl', coverage = 0.50, ...)
global_envelope_test(curve_sets, type = 'erl', alpha = 0.05, ...)
\end{Code}
where the multivariate or functional data are provided in \code{curve_sets},
\code{type} specifies type of the global envelope (see Table~\ref{tab:GEoverview} and descriptions in Appendix~\ref{app:GEdef}), and the coverage or level of the global envelope is specified by \code{coverage} or \code{alpha} ($=1-$\code{coverage}), respectively.
Additionally, one can, for example, specify the one or two-sided \code{alternative}, i.e., whether only small or large values of $\TT_i$ or both should be considered extreme.
These two functions are the core functions for global envelopes in the package \pkg{GET}: given an appropriate set of curves, or, in fact vectors, they can be used for producing global envelopes of Table~\ref{tab:GEoverview} in all tasks (i)-(iii) listed in Section~\ref{sec:motivating_example}.

Recently, the argument \code{typeone} was added to \fct{global\_envelope\_test}. This argument can be used to specify the control for the global test, FWER or FDR, where the former (default) leads to the tests of Table~\ref{tab:GEoverview}, as further specified by the argument \code{type}, and the latter to the FDR envelopes proposed in \citet{MrkvickaMyllymaki2023}.

Different objects are supported for the data in \code{curve_sets} (see help files of the functions and examples below),
but the basic form provided by the \pkg{GET} package is a \class{curve\_set} object that can be constructed by the function \fct{curve\_set} simply providing the observed and/or simulated curves, and optionally the (one- or two-dimensional) argument values where the curves have been observed (see Section~\ref{sec:GOT} for an example). The function \fct{curve\_set} takes care of checking the appropriateness of the data, and saving the data in the form that contains the relevant information of the curves for global envelope methods, in particular for plotting the results with graphical interpretation (see examples in Section~\ref{sec:illustrations}).

In addition to constructing global envelopes from a set of curves, the \fct{central\_region} and \fct{global\_envelope\_test} functions provide combined central regions or combined global envelope tests if the user provides a list consisting of (appropriate) sets of curves in the argument \code{curve_sets}.
Details of the combining methodology is given in Appendix~\ref{app:combinedGE} and examples are given in Sections~\ref{sec:motivating_example} and \ref{sec:illustrations}.
The \pkg{GET} package also provides functions for specific tasks (see Table~\ref{tab:FUNCTIONSoverview} and the examples in Section~\ref{sec:illustrations}). These functions utilize \fct{central\_region} and \fct{global\_envelope\_test} for the global envelope construction. In addition, many of these functions take care of preparing the simulations or permutations for the specific testing task.
The \fct{print} and \fct{plot} methods are available for the objects obtained as
the output of the global envelope methods of \pkg{GET}. The plots present the results with IGI.
They are produced using the \pkg{ggplot2} package \citep{Wickham2016}.

\begin{table}[t!]
\centering
\begin{tabular}{lp{10.0cm}}
\hline
Function name              & Description \\ \hline
\fct{curve\_set} & Create a \class{curve\_set} out of data given in the right form \\
\fct{crop\_curves} & Crop curves \\
\fct{forder} & Different measures for ordering the multivariate statistics from the most extreme to least extreme one \\
\fct{central\_region} & Central regions or global envelopes or confidence bands with IGI (see types in Table~\ref{tab:GEoverview}) \\
\fct{global\_envelope\_test} & Global envelope tests \\ 
\fct{GET.composite}          & Adjusted global envelope tests for composite null hypotheses \\ 
\fct{fBoxplot} & Functional boxplot based on a central region with IGI \\
\fct{graph.fanova}  & One-way ANOVA tests for functional data with graphical interpretation \citep{MrkvickaEtal2020} \\
\fct{frank.fanova}  & One-way functional ANOVA tests based on the global envelopes applied to $F$~values \citep{MrkvickaEtal2020} \\
\fct{graph.flm}  &  Non-parametric graphical tests of significance in functional general linear model (GLM) \citep{MrkvickaEtal2021a} \\
\fct{frank.flm} & Global envelope tests applied to $F$~values in permutation inference for the GLM \citep{MrkvickaEtal2022} \\
\fct{fclustering} & Functional clustering based on a specified measure \citep{DaiEtal2022} \\
\fct{GET.distrequal} & Graphical $n$ sample test of correspondence of distribution functions \\
\fct{GET.distrindep} & Permutation-based tests of independence to samples from any bivariate distribution \citep{DvorakMrkvicka2022} \\
\fct{GET.spatialF} & Testing global and local covariate effects in point process models \citep{MyllymakiEtal2020} \\
\end{tabular}
\caption{\label{tab:FUNCTIONSoverview} Key and special purpose functions in the \pkg{GET} package. More complete list of functions can be found from the main help page of the package available by typing \code{help('GET-package')} in \proglang{R}.}
\end{table}


\section[Using GET]{Using \pkg{GET}} \label{sec:illustrations}

\subsection{Sets of functions, ordering, combining, and central regions}\label{sec:cr}

\begin{figure}[ht!]
\centering
\includegraphics{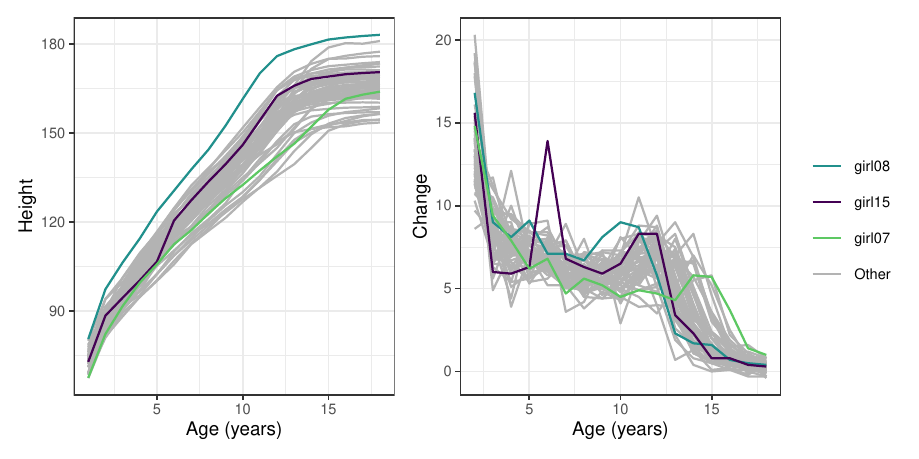}
  \caption{The heights (left) and height differences (right) of the 54 girls of the \code{growth} data of the \proglang{R} package \pkg{fda} at ages from 1 to 18. Three girls having the most extreme curves (combined ordering by the area measure) are highlighted with the colors specified in the legend.}
  \label{fig:girls_curves}
\end{figure}

The \proglang{R} package \pkg{fda} contains Berkeley Growth Study data \citep{RamsaySilverman2005} of the heights of 39 boys and 54 girls from ages 1 to 18 and the ages at which the data were collected.
To illustrate different features of \pkg{GET}, we investigated whether there are any outliers in the girls regarding both their annual heights and changes within years.
First two \class{curve\_set} objects were created containing the raw heights and the differences within the years (see Figure~\ref{fig:girls_curves}):
\begin{Schunk}
\begin{Sinput}
R> library("fda")
R> years <- paste(1:18)
R> curves <- growth[['hgtf']][years,]
R> cset1 <- curve_set(r = as.numeric(years), obs = curves)
R> cset2 <- curve_set(r = as.numeric(years[-1]),
+    obs = curves[-1,] - curves[-nrow(curves),])
\end{Sinput}
\end{Schunk}

Ordering the functions from the most extreme to the least extreme by the \code{'area'} measure (see Table~\ref{tab:GEoverview} and Appendix~\ref{app:GEdef}), the 8th girl was observed to have the most extreme heights and the 15th girl the most extreme changes (below the first ten most extreme girl indices are printed):
\begin{Schunk}
\begin{Sinput}
R> A1 <- forder(cset1, measure = 'area'); order(A1)[1:10]
\end{Sinput}
\begin{Soutput}
 [1]  8 13 29 48 42 25  7 38 18 40
\end{Soutput}
\begin{Sinput}
R> A2 <- forder(cset2, measure = 'area'); order(A2)[1:10]
\end{Sinput}
\begin{Soutput}
 [1] 15  7  3  8 25 52 19 16 24  5
\end{Soutput}
\end{Schunk}
Generally, ordering with respect to heights or height differences leads to two different orderings of the girls. Combined ordering can be done by combining these two by the ERL measure as described in Appendix~\ref{app:combinedGE} (two-step combining procedure). In \proglang{R}, the two sets of curves need to be provided in a list to the function \fct{forder}:
\begin{Schunk}
\begin{Sinput}
R> csets <- list(Height = cset1, Change = cset2)
R> A <- forder(csets, measure = 'area'); order(A)[1:10]
\end{Sinput}
\begin{Soutput}
 [1]  8 15  7 13  3 29 48 25 42 52
\end{Soutput}
\end{Schunk}
Figure~\ref{fig:girls_curves} highlights the curves of the three most extreme girls.
The plots of the two sets of curves were produced using the \pkg{GET} and \pkg{ggplot2} packages and combined by the \pkg{patchwork} package \citep{patchwork2020}:
\begin{Schunk}
\begin{Sinput}
R> library("ggplot2")
R> library("patchwork")
R> cols <- c("#21908CFF", "#440154FF", "#5DC863FF")
R> p1 <- plot(cset1, idx = order(A)[1:3], col_idx = cols) +
+    labs(x = "Age (years)", y = "Height")
R> p2 <- plot(cset2, idx = order(A)[1:3], col_idx = cols) +
+    labs(x = "Age (years)", y = "Change")
R> p1 + p2 + plot_layout(guides = "collect")
\end{Sinput}
\end{Schunk}

The labels were above redefined for the default plots by the function \fct{labs} of the \pkg{ggplot2} package. In general, parts of the default plots of \pkg{GET} can be edited in this manner using \pkg{ggplot2} functions such as \fct{labs} and \fct{ggtitle}.

\begin{figure}[ht!]
\centering
\includegraphics{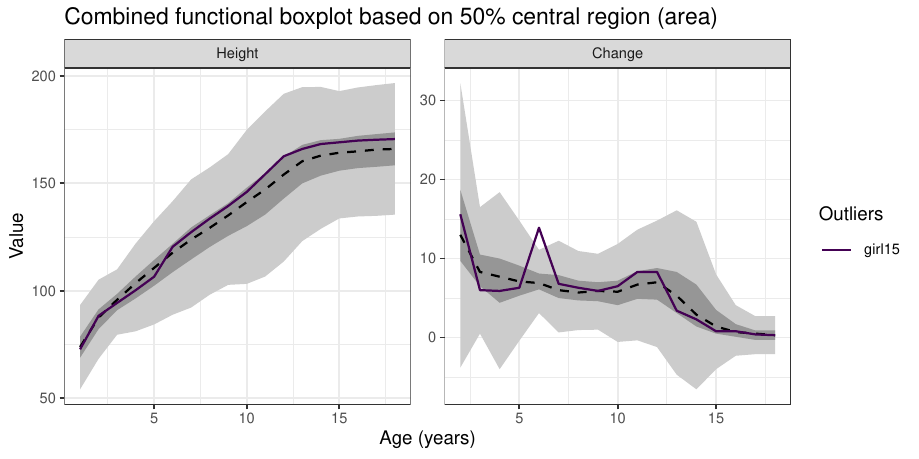}
\caption{The functional boxplot (entire gray band) using the 50\% central region (inner dark gray band) and the expansion factor 1.5 jointly for the heights and changes of heights of the 54 girls (see Figure~\ref{fig:girls_curves}). The solid line is an observed data function (vector) that goes outside the functional boxplot (the 15th girl in the data).}
  \label{fig:girls_fboxplot}
\end{figure}

The combined central region can be constructed directly by passing the sets of curves to the function \fct{central\_region}, in a similar manner as for \fct{forder} above.
By using the functional boxplot \citep{SunGenton2011} with the same measure and central region, an investigation can be made into whether the most extreme girls are outliers with respect to height or its change. Figure~\ref{fig:girls_fboxplot} shows the 50\% central region and the functional boxplot with the inflation factor 1.5 jointly for the heights and their changes obtained by:
\begin{Schunk}
\begin{Sinput}
R> res <- fBoxplot(csets, type = 'area', factor = 1.5)
R> plot(res) + labs(x = "Age (years)", y = "Value")
\end{Sinput}
\end{Schunk}
One can see that one of the girls (the 15th girl) is an outlier, because she has grown extraordinarily much in her sixth year. However, the highest height curve of Figure~\ref{fig:girls_curves} (left) is not regarded as an outlier with the given specifications.

Note that the combined central region computed using any measure of Table~\ref{tab:GEoverview} has IGI.
On the contrary, central regions computed with the use of band depths implemented in the \pkg{fda} package do not satisfy IGI. \citet{NarisettyNair2016} proposed central regions and functional boxplots based on the ERL measure (see Table~\ref{tab:GEoverview}) and compared them to those based on band depths, claiming more reasonable behavior.
Note also that IGI is not a property of the functional boxplot.

\subsection{Monte Carlo goodness-of-fit testing for simple hypotheses: The example of testing complete spatial randomness}\label{sec:GOT}

\begin{figure}[htb]
\centering
\includegraphics{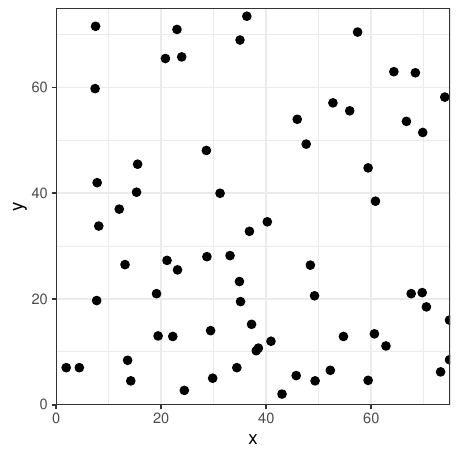}
  \caption{Locations of 67 trees with height $> 25$ m observed in an area of $75\text{ m}\times 75\text{ m}$.}
  \label{fig:trees}
\end{figure}

Figure~\ref{fig:trees} shows the locations of 67 large trees (with height $> 25$ m) in an area of size $75\text{ m}\times 75\text{ m}$ from an uneven aged multi-species broadleaf nonmanaged forest in Kaluzhskie Zaseki, Russia \citep{GrabarnikChiu2002, Lieshout2010}. The $x$- and $y$-coordinates of the locations are available in the data \code{adult_trees} in the \pkg{GET} package.

The test of complete spatial randomness (CSR) is a typical first step in analyzing a spatial point pattern such as the tree pattern of Figure~\ref{fig:trees}. CSR along with other hypotheses for spatial point patterns are commonly tested using an estimator of a summary function that is a function of distance $r$, e.g., Ripley's $K$ function or its transformation $L(r)=\sqrt{K(r)/\pi}-r$ for $r\geq 0$ \citep{Ripley1977, Besag1977}. In this context, one typically resorts to the Monte Carlo simulation \citep[see, e.g.,][]{IllianEtal2008, Diggle2013a, MyllymakiEtal2017}.
First, this example is used to show the general steps to prepare a global envelope test for testing a simple hypothesis.
Second, it is shown how the same example of testing a simple hypothesis for a spatial point pattern can be performed by utilizing the \proglang{R} package \pkg{spatstat}.

The testing of a simple hypothesis does not require the estimation of any model parameters, and the one-stage test illustrated below can be used. 
In the case of a composite null hypothesis, the level of the test needs to be adjusted, see Section~\ref{sec:GOTcomposite} and Appendix~\ref{app:adjustedGE}.

\subsubsection{General setup for testing simple hypotheses}

The first step of a Monte Carlo test is to generate \code{nsim} simulations under the null hypothesis and to calculate the chosen test function (vector) for the data and simulations. 
Here the functions \fct{runifpoint} and \fct{Lest} of \pkg{spatstat} are used to generate a simulation from the binomial process (CSR with the number of points fixed to the observed number of points in the pattern \code{X}) and to estimate the centred $L$-function for a pattern, respectively:
\begin{Schunk}
\begin{Sinput}
R> library("spatstat.explore")
R> data("adult_trees")
R> X <- as.ppp(adult_trees, W = square(75))
R> nsim <- 999
R> obs.L <- Lest(X, correction = "translate")
R> r <- obs.L[['r']]
R> obs <- obs.L[['trans']] - r
R> sim <- matrix(nrow = length(r), ncol = nsim)
R> for(i in 1:nsim) {
+    sim.X <- runifpoint(ex = X)
+    sim[, i] <- Lest(sim.X, correction = "translate", r = r)[['trans']] - r
+  }
\end{Sinput}
\end{Schunk}
Thereafter, the function \fct{curve\_set} can be used to construct a \class{curve\_set} object from the argument values where the test vectors were evaluated (\code{r}), the observed vector (\code{obs}) and the simulated vectors (\code{sim}):
\begin{Schunk}
\begin{Sinput}
R> cset <- curve_set(r = r, obs = obs, sim = sim)
\end{Sinput}
\end{Schunk}
In some cases, missing or infinite values can occur for the computed statistic at some chosen $r$ (e.g., for too large $r$ with another spatial summary function $J$). These problematic, uninteresting $r$-values can be easily omitted from \code{cset} using the function \fct{crop\_curves}.
The final step is to make the global envelope test on the given set of vectors:
\begin{Schunk}
\begin{Sinput}
R> res <- global_envelope_test(cset, type = 'erl')
R> plot(res) + ylab(expression(italic(hat(L)(r)-r)))
\end{Sinput}
\end{Schunk}
\begin{figure}[htb]
\centering
\includegraphics{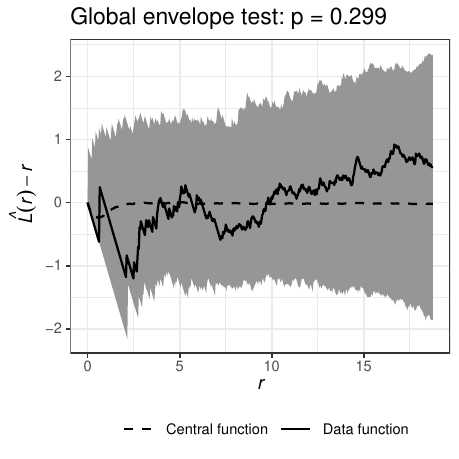}
  \caption{The global envelope test for the CSR of the tree pattern of Figure~\ref{fig:trees} using the centred $L$-function. The gray band represents the 95\% global envelope (\code{'erl'}).}
  \label{fig:trees_CSR}
\end{figure}
In this manner, the global envelope test can be constructed for any simple hypothesis and any test vector, as long as one can generate the required simulations and calculate the test vectors.

The test output is shown in Figure~\ref{fig:trees_CSR} (left), which shows no evidence against CSR \citep[see more detailed description in][Section~S4]{MyllymakiEtal2017}.

\subsubsection[Testing simple hypotheses for a point pattern utilizing the R package spatstat]{Testing simple hypotheses for a point pattern utilizing the \proglang{R} package \pkg{spatstat}}

For point process testing, the \pkg{GET} package and \fct{global\_envelope\_test} support the use of the \proglang{R} package \pkg{spatstat} \citep{spatstat2015} for the simulations and calculations of the summary functions by the function \fct{envelope}. Namely, the object returned by \fct{envelope} can simply be given to the function \fct{global\_envelope\_test} in the argument \code{curve_sets}. Importantly, the functions must be saved setting \code{savefuns = TRUE} in the \fct{envelope} call:
\begin{Schunk}
\begin{Sinput}
R> env <- envelope(X, nsim = 999, fun = "Lest", correction = "translate",
+    transform = expression(.-r), simulate = expression(runifpoint(ex = X)),
+    savefuns = TRUE, verbose = FALSE)
R> res <- global_envelope_test(env, type = 'erl')
\end{Sinput}
\end{Schunk}
Above the arguments \code{fun}, \code{correction} and \code{transform} define the summary function to be calculated (the latter two parameters are passed to the function \fct{Lest}) and \code{simulate} specifies how the patterns are simulated under the null hypothesis (here CSR). The result can be plotted similarly as above.

Further examples of use of the \pkg{GET} package for point pattern analysis are given in an accompanying vignette available in \proglang{R} by typing \code{library("GET")} and \code{vignette("pointpatterns")}.

\subsection{Monte Carlo goodness-of-fit testing for composite hypotheses: The example of graphical normality test}\label{sec:GOTcomposite}

\cite{AldorNoimanEtal2013} provided a graphical test for normality for simple hypotheses (i.e., known parameters of sample distribution) based on a qq-plot envelope, whose shape was derived from theoretical properties of quantiles of the uniform distribution. They also provided a version of this algorithm for composite hypotheses (i.e., unknown parameters of sample distribution). However, according to our unpublished study, this test does not achieve the required significance level. Therefore, the example of the exact adjustment for the composite hypothesis is provided here, based on the two-stage procedure of \cite{BaddeleyEtal2017} (see Appendix~\ref{app:adjustedGE}). For simplicity, in this example, this adjustment is applied directly to the empirical distribution functions. Apparently, the adjustment could also be applied to the qq-plot envelopes of \cite{AldorNoimanEtal2013}.

The normality test is illustrated for nitrogen oxides (NO$_x$) emission levels available in the data \code{poblenou} from the \proglang{R} package \pkg{fda.usc} \citep{fdausc2012}. The data contains NO$_x$ emission levels ($\mu$g/m$^3$) measured every hour by a control station close to an industrial area in Poblenou in Barcelona (Spain) for 115 days from 23 February to 26 June, 2005. NO$_x$ is a pollutant which is caused by combustion processes in sources that burn fuels, e.g., motor vehicles, electric utilities, and industries \citep{FebreroEtal2008}.
In Section~\ref{sec:fANOVA}, the whole functional trajectories of 24 h observations are studied, but for illustration purposes, here the attention is restricted to the NO$_x$ levels at 10 am. 

A general solution to make the adjusted test is to prepare all the required simulations and provide them to the function \fct{GET.composite} in arguments \code{X} and \code{X.ls}. Let \code{dat} be a vector containing the data values and $n$ is the number of observations specified as follows:
\begin{Schunk}
\begin{Sinput}
R> library("fda.usc")
R> data("poblenou")
R> dat <- poblenou[['nox']][['data']][,'H10']
R> n <- length(dat)
\end{Sinput}
\end{Schunk}
Then, first, the parameters of the normal distribution are estimated (1.\ step of the algorithm of Appendix~\ref{app:adjustedGE})
\begin{Schunk}
\begin{Sinput}
R> mu <- mean(dat)
R> sigma <- sd(dat)
\end{Sinput}
\end{Schunk}
and, using the function \fct{ecdf} of the \proglang{R} package \pkg{stats} \citep{R2020}, the empirical cumulative distribution functions are calculated for the data and for \code{nsimsub} replicates of $n$ simulations from the fitted normal distribution (2.\ step):
\begin{Schunk}
\begin{Sinput}
R> nsim <- nsimsub <- 199 # The number of simulations
R> r <- seq(min(dat), max(dat), length = 100)
R> obs <- stats::ecdf(dat)(r)
R> sim <- replicate(nsimsub, {
+    x <- rnorm(n, mean = mu, sd = sigma)
+    stats::ecdf(x)(r)
+  })
R> cset <- curve_set(r = r, obs = obs, sim = sim)
\end{Sinput}
\end{Schunk}
Here the last command creates a \class{curve\_set} object of the observed and simulated empirical cumulative distribution functions.
Thereafter, another \code{nsim} replicates of the $n$ simulations from the fitted model are simulated, and the same calculations as above for the data are done for each of these simulations (steps 3.-4.\ of the algorithm of Appendix~\ref{app:adjustedGE}): 
\begin{Schunk}
\begin{Sinput}
R> cset.ls <- list()
R> for(rep in 1:nsim) {
+    x <- rnorm(n, mean = mu, sd = sigma)
+    mu2 <- mean(x)
+    sigma2 <- sd(x)
+    obs2 <- stats::ecdf(x)(r)
+    sim2 <- replicate(nsimsub, {
+      x2 <- rnorm(n, mean = mu2, sd = sigma2)
+      stats::ecdf(x2)(r)
+    })
+    cset.ls[[rep]] <- curve_set(r = r, obs = obs2,
+      sim = sim2)
+  }
\end{Sinput}
\end{Schunk}
Thus, the list \code{cset.ls} contains all the simulations from the second stage of the algorithm. As a final step, \fct{GET.composite} can be used to prepare the adjusted test:
\begin{Schunk}
\begin{Sinput}
R> res <- GET.composite(X = cset, X.ls = cset.ls, type = 'erl')
R> plot(res) + labs(x = "NOx", y = "Ecdf")
\end{Sinput}
\end{Schunk}
%
%

\begin{figure}[ht!]
  \centering
\includegraphics{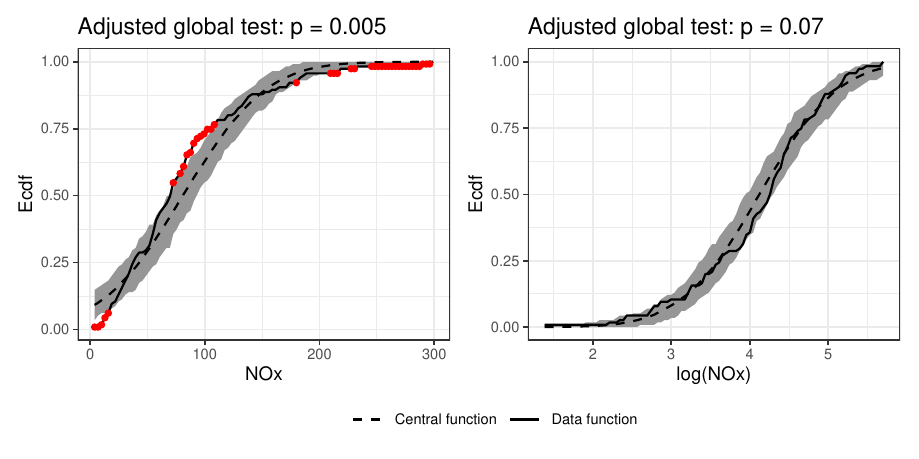}
  \caption{Graphical normality test for the NO$_x$ (left) and logarithm of the NO$_x$ (right) levels at 10 am.
  The gray band represents the 95\% global envelope (\code{'erl'}). Red dots are attached to the data function outside the envelope.}
  \label{fig:normalitytest}
\end{figure}

Figure~\ref{fig:normalitytest} (left) shows the test result for the NO$_x$ levels at 10 am. One can see that the normality does not hold according to the test: the estimated distribution function is skewed to the right with respect to the normal envelope. Therefore, we further applied the same normality test to the logarithm of the NO$_x$ values as well, and then the normality hypothesis was not rejected (Figure~\ref{fig:normalitytest}, right).

\subsection{Graphical $n$-sample test of correspondence of distribution functions}\label{sec:CDF}

\begin{figure}[ht!]
  \centering
\includegraphics{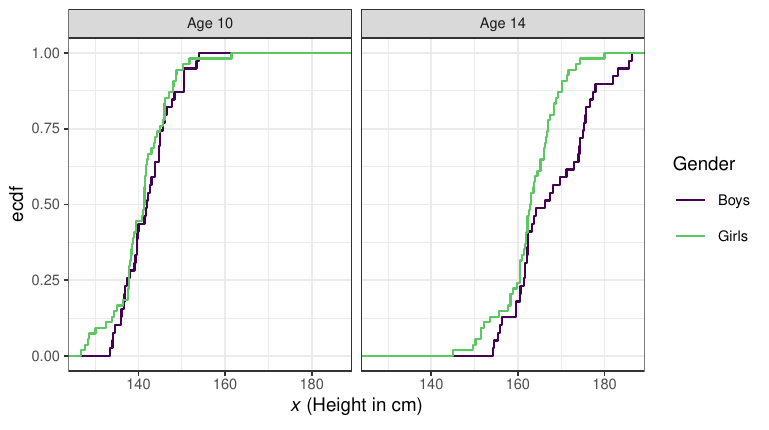}
  \caption{The empirical cumulative distribution functions of the heights of the 54 girls and 39 boys of the \code{growth} data of the \proglang{R} package \pkg{fda} at ages 10 (left) and 14 (right).}
  \label{fig:necdfs}
\end{figure}

\begin{figure}[ht!]
  \centering
\includegraphics{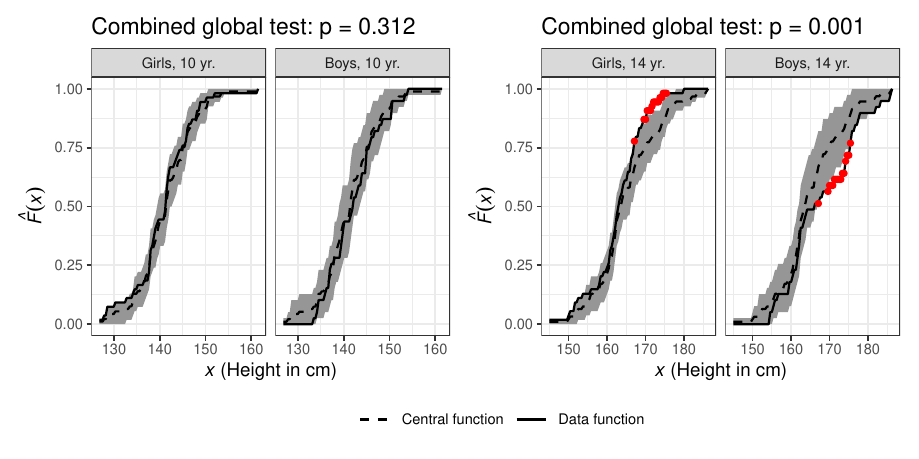}
  \caption{Global envelope tests for comparison of the empirical cumulative distribution functions of the heights of the girls and boys (see Figure~\ref{fig:necdfs}). Red color indicates the heights where the observed distribution functions go outside the 95\% global envelope (\code{'erl'}; gray bands). Left: Age 10; Right: Age 14.}
  \label{fig:necdfs_means_GET}
\end{figure}

The graphical $n$-sample test of correspondence of distribution functions serves as a simple example of permutation tests.
Figure~\ref{fig:necdfs} shows the empirical cumulative distribution functions (ECDFs) obtained by the function \fct{ecdf} of the \proglang{R} package \pkg{stats} \citep{R2020} for the heights of the 54 girls and 39 boys of the \code{growth} data \citep[see above, and][]{RamsaySilverman2005} at ages 10 (left) and 14 (right).
A global envelope test can be performed to investigate whether the two (or more generally $n$) distribution functions differ from each other significantly and how they differ. This test is a generalization of the two-sample Kolmogorov-Smirnov test with a graphical interpretation. The generalization is provided in two ways, in the number of samples and in the variable width of the envelopes. Namely, the Kolmogorov-Smirnov test provides an envelope of constant width which suffers in determining differences in extreme quantiles.
Here it is assumed that the heights in the sample $i$ are an i.i.d.\ sample from the distribution $F_i(r)$, $i=1, \ldots, n$, and the hypothesis $F_1(r)=\ldots = F_n(r)$ is to be tested.
The simulations under the null hypothesis that the distributions are the same can be obtained by permuting the individuals of the groups. The \pkg{GET} package provides the wrapper function \fct{GET.distrequal} that can be used to compare $n$ distribution functions graphically, $n=2,3,\ldots$. The (default) test vector is
\begin{equation*}
\mathbf{T} = (\hat{F}_1(r), \ldots, \hat{F}_n(r)),
\end{equation*}
where $\hat{F}_i(r) = (\hat{F}_i(r_1), \ldots, \hat{F}_i(r_d))$ is the ECDF of the $i$th sample evaluated at argument values $r = (r_1,\ldots,r_d)$.
To test the equality of distributions, one simply needs to provide the samples as a list (code for age 10 shown here) for \fct{GET.distrequal} and plot the object returned by \fct{GET.distrequal} (Figure~\ref{fig:necdfs_means_GET}, left):
\begin{Schunk}
\begin{Sinput}
R> fm10.l <- list('Girls, 10 yr.' = growth$hgtf["10",],
+                 'Boys, 10 yr.' = growth$hgtm["10",])
R> res10 <- GET.distrequal(fm10.l, nsim = 1999)
R> myxlab <- substitute(paste(italic(i), " (", j, ")", sep = ""),
+      list(i = "x", j = "Height in cm"))
R> plot(res10) + xlab(myxlab)
\end{Sinput}
\end{Schunk}
The height distributions at age 10 do not differ from each other significantly, but at age 14 the boys are taller, particularly with a difference that the proportion of girls reaching a height of around 175 cm is much lower (Figure~\ref{fig:necdfs_means_GET}, right).

\subsection{Graphical functional one-way ANOVA}\label{sec:fANOVA}

\begin{figure}[ht!]
  \centering
\includegraphics{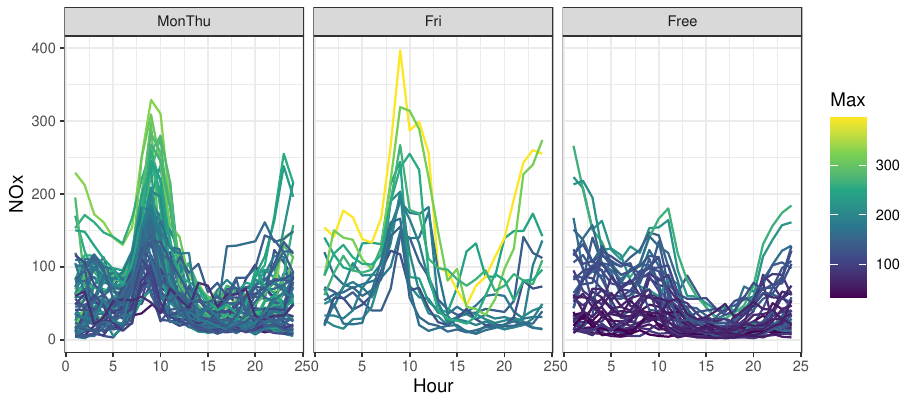}
\caption{The NO$_x$ levels for Monday-Thursday (Mon-Thu), Friday (Fri) and non-working days (Free) including weekend and festive days in Poblenou for 115 days from 23 February to 26 June, 2005. The color of the daily curves is according to the maximum NO$_x$ level ($\mu$g/m$^3$) of the day.}
  \label{fig:NOx_data}
\end{figure}

The use of the function \fct{graph.fanova} of the \pkg{GET} package for the graphical functional one-way ANOVA is illustrated using the data set \code{poblenou} of the \proglang{R} package \pkg{fda.usc} \citep[][see also Section~\ref{sec:GOTcomposite} above]{fdausc2012}.
The trajectories of the 24 h observations of the NO$_x$ levels for Monday-Thursday (MonThu), Friday (Fri) and non-working days (Free) including weekend and festive days (Figure~\ref{fig:NOx_data}) are compared. For the purposes of this example, a factor vector \code{Type} was prepared containing the type of the day for each of the 115 days having levels \code{"MonThu"}, \code{"Fri"} and \code{"Free"}:
\begin{Schunk}
\begin{Sinput}
R> library("fda.usc")
R> data("poblenou")
R> fest <- poblenou$df$day.festive; week <- as.integer(poblenou$df$day.week)
R> Type <- vector(length = length(fest))
R> Type[fest == 1 | week >= 6] <- "Free"
R> Type[fest == 0 & week %in% 1:4] <- "MonThu"
R> Type[fest == 0 & week == 5] <- "Fri"
R> Type <- factor(Type, levels = c("MonThu", "Fri", "Free"))
\end{Sinput}
\end{Schunk}

Assuming that the NO$_x$ levels $T_{ij}(r)$ at times $r \in R = [0,24]$ are i.i.d.\ samples from stochastic processes $SP(\mu_j, \gamma_j)$ with mean functions $\mu_j(r)$, $r\in R$, and covariance functions $\gamma_j(s, t)$, $s, t \in R$, for $j=1, \ldots , J$ (here $J=3$), the groups of NO$_x$ levels can be compared by means of the graphical functional ANOVA \citep{MrkvickaEtal2020}. The hypothesis 
\begin{equation*}
H_0 : \mu_1(r) = \ldots = \mu_J(r), r \in R,
\end{equation*}
can be tested by the test statistic 
\begin{equation}\label{eq:graphfanova_means}
\TT = (\overline{T}_{1}, \overline{T}_{2}, \ldots , \overline{T}_{J}),
\end{equation}
where $\overline{T}_{j} = (\overline{T}_{j}(r_1), \ldots , \overline{T}_{j}(r_d))$ is the mean of functions in the $j$th group at the discrete number of arguments $r_1,\ldots,r_d$ (here each hour of the day).
The hypothesis can be equivalently expressed as
\begin{equation*}
H'_0 : \mu_{j'}(r) - \mu_j(r) = 0, r \in R,  j' =1, \ldots , J-1, j =j', \ldots , J
\end{equation*}
and an alternative test statistic is
\begin{equation}\label{eq:graphfanova_contrasts}
\TT' = (\overline{T}_{1}-\overline{T}_{2}, \overline{T}_{1}-\overline{T}_{3}, \ldots , \overline{T}_{J-1}-\overline{T}_{J}),
\end{equation}
where we used the notation $\overline{T}_{j}-\overline{T}_{j'} = (\overline{T}_{j}(r_1)-\overline{T}_{j'}(r_1), \ldots , \overline{T}_{j}(r_d)-\overline{T}_{j'}(r_d))$
The latter test statistic (Equation~\ref{eq:graphfanova_contrasts}) can be obtained by setting \code{contrasts = TRUE} in the call of \fct{graph.fanova}.

\begin{figure}[ht!]
  \centering
\includegraphics{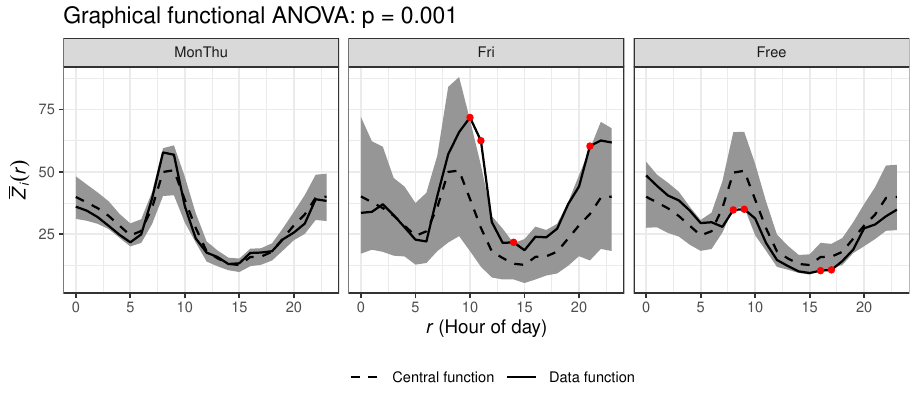}
  \caption{The test for the equality of variances of the NO$_x$ levels observed each hour ($r$) of the day. The 95\% global envelope (\code{'erl'}; gray band) representing the null hypothesis of equal variances and the observed test statistic, the mean of functions $Z_{ij}(r) = | T_{ij}(r) - \overline{T}_j(r) |$ in each group (solid line with red dots when outside the envelope).}
  \label{fig:graphfanova_NOx_var_plot}
\end{figure}

Because the ANOVA design assumes equal covariance functions, we first tested for the homoscedasticity using the test proposed by \citet[][Section~2.3]{MrkvickaEtal2020} and implemented in \pkg{GET}. We created the \class{curve\_set} object
\begin{Schunk}
\begin{Sinput}
R> cset <- curve_set(obs = t(poblenou[['nox']][['data']]), r = 0:23)
\end{Sinput}
\end{Schunk}
and we used the function \fct{graph.fanova} with the option \code{test.equality = "var"} to test the equality of variances (Figure~\ref{fig:graphfanova_NOx_var_plot}):
\begin{Schunk}
\begin{Sinput}
R> res.v <- graph.fanova(nsim = 2999, curve_set = cset, groups = Type,
+                        test.equality = "var")
R> myxlab <- substitute(paste(italic(i), " (", j, ")", sep = ""),
+      list(i = "r", j = "Hour of day"))
R> plot(res.v) + xlab(myxlab)
\end{Sinput}
\end{Schunk}
Here the test suggests heteroscedasticity and also \citet{FebreroEtal2008} assumed heteroscedasticity of working and non-working days.
Therefore, we applied correction for unequal variances to the three groups by rescaling the functions $T_{ij}(r)$ of $J$ groups containing $n_1, \ldots , n_J$ functions observed on the finite interval $R=[0, 24]$ by the transformation
\begin{equation}\label{eq:scaling}
Y_{ij}(r) = \frac{T_{ij}(r) - \overline{T}_j(r)}{\sqrt{\text{Var}(T_j(r))}}\cdot \sqrt{\text{Var}(T(r))} + \overline{T}_j(r), \quad j = 1, \ldots, J,\; i = 1, \ldots, n_j, 
\end{equation}
where the group sample mean $\overline{T}_j(r)$ and overall sample variance $\text{Var}(T(r))$ are involved
to keep the mean and variability of the functions at the original scale. The group sample variance $\text{Var}(T_j(r))$ corrects the unequal variances. 
This scaling is applied to the set of curves given to the function \fct{graph.fanova} if the user specifies \code{variances = "unequal"} (the default is no correction, \code{variances = "equal"}).
When using the rescaled functions, the test vectors (Equations~\ref{eq:graphfanova_means} and \ref{eq:graphfanova_contrasts}) are asymptotically exchangeable under the null hypothesis of equal means only under the assumption of normality of stochastic processes $SP(\mu_j,\gamma_j)$ \citep{MrkvickaEtal2020}. Therefore, the log transformation was applied to the NO$_x$ values prior to the transformation (Equation~\ref{eq:scaling}):
\begin{Schunk}
\begin{Sinput}
R> lcset <- curve_set(obs = t(log(poblenou[['nox']][['data']])), r = 0:23)
\end{Sinput}
\end{Schunk}

\begin{figure}[ht!]
  \centering
\includegraphics{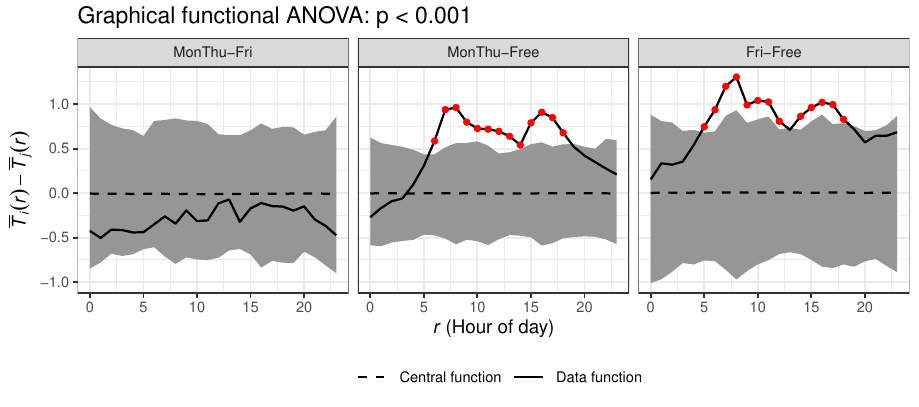}
  \caption{The output of the graphical functional ANOVA to test the difference between the type of the day on the log NO$_x$ levels observed each hour ($r$) of the day. The 95\% global envelope (\code{'erl'}; gray band) accompanied with the observed differences between the group means (solid line with red dots when outside the envelope).}
  \label{fig:graphfanova_logNOx}
\end{figure}

To sample from the null hypotheses, the simple permutation of raw functions among the groups is performed. The permutations and the global envelope test can be done by the \fct{graph.fanova} function (Figure~\ref{fig:graphfanova_logNOx}):
\begin{Schunk}
\begin{Sinput}
R> res.c <- graph.fanova(nsim = 2999, curve_set = lcset, groups = Type,
+    variances = "unequal", contrasts = TRUE)
R> plot(res.c) + xlab(myxlab)
\end{Sinput}
\end{Schunk}

Thus, the test rejects the null hypothesis $H'_0$ that the differences between the groups would be zero and shows that on Monday-Thursday and Friday the (log) NO$_x$ levels are significantly higher than on free days basically during the whole day with peaks around 8 am and 4 pm. The difference between Monday-Thursday and Friday was not significant.

The graphical functional ANOVA allows one to detect either a) which groups deviate from the mean (default) or b) which specific groups are different (option \code{contrasts = TRUE}). The example above was for the latter. Note that this test directly has the nature of a post hoc test. Furthermore, both versions of the test allow one to identify which $r$ values lead to the potential rejection of the null hypothesis.

When a graphical interpretation for group specific differences is not of interest but the area of rejection is, instead of \fct{graph.fanova} it is possible to apply the one-way functional ANOVA based on the $r$-wise $F$~statistics, $r\in R$. This test is implemented in the function \fct{frank.fanova}. For the log NO$_x$ data, the test result was that there are differences between the groups for the hours from 5 am to 6 pm (figure omitted).

\subsection{Functional GLM and two-dimensional plots}\label{sec:fGLM}

\begin{figure}[ht!]
\centering
\includegraphics{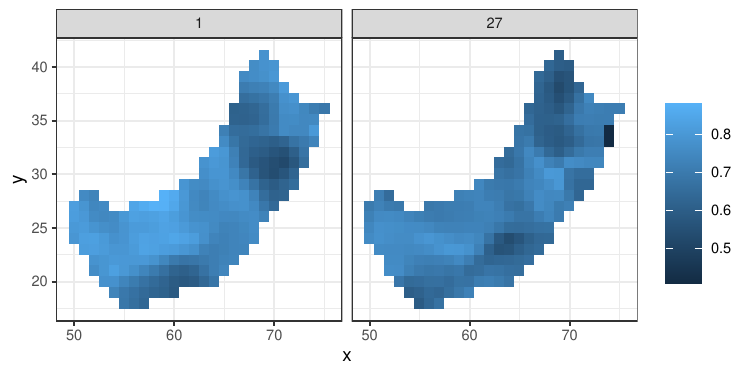}
  \caption{The example brain image data in the right Crus Cerebellum 1 region at a slice for one subject from the autism group (subject number 1) and one from the control group (subject number 27).}
  \label{fig:abide_9002_23_subj1and27}
\end{figure}

Similar type of methods as in the functional one-way ANOVA above can be used in a more general setup of functional general linear models (GLMs).
The global envelopes for functional GLMs are illustrated here by an example of a small subset of the autism brain imaging data collected by resting state functional magnetic resonance imaging (R-fMRI) \citep{DiMartinoEtal2014}. The preprocessed fMRI data contains measurements from 514 individuals with the autism spectrum disorder (ASD) and 557 typical controls (TC), where subjects with low quality on imaging data or having a large proportion of the missing values were removed. The imaging measurement for local brain activity at resting state was fractional amplitude of low frequency fluctuations \citep{ZouEtal2008}.
The data considered here and available as the data object \code{abide_9002_23} in the \pkg{GET} package contains data from one of the 116 different anatomical regions in the brain partitioning being based on the anotomical automatic labeling system of \citet{Tzourio-MazoyerEtal2002}.
The studied region is the right Crus Cerebellum 1 region of the brain at one slice (23) accompanied with three subject-specific factors, i.e., group (autism and control), sex and age. Figure~\ref{fig:abide_9002_23_subj1and27} obtained by
\begin{Schunk}
\begin{Sinput}
R> data("abide_9002_23")
R> plot(abide_9002_23[['curve_set']], idx = c(1, 27))
\end{Sinput}
\end{Schunk}
shows the data for two subjects, illustrating the small region used as the example.
In the examples below, the effect of the group on the images is studied in the presence of nuisance factors sex and age.

\subsubsection{Graphical functional GLM}

The functional GLM is the general linear model
\begin{equation}\label{eq:GLM}
\mathbf Y(r) = \mathbf X(r) \Bbeta(r) + \mathbf Z(r) \Bgamma(r) + \epsilon(r)
\end{equation}
where the argument $r \in \{1, \ldots, d\}$ determines the component of the vector or the spatial point or pixel of an image. 
For every argument $r$, a one-dimensional GLM is considered with $\mathbf X(r)$ being a $n \times k$ matrix of regressors of interest (here group), $\mathbf Z(r)$ being a $n \times l$ matrix of nuisance regressors (here sex and age), $\mathbf Y(r)$ being a $n \times 1$ vector of observed data, and $\epsilon(r)$ being a $n \times 1$ vector of random errors with a mean of zero and a finite variance $\sigma^2(r)$ for every $r \in I$. Further, $\Bbeta(r)$ and $\Bgamma(r)$ are the regression coefficient vectors of dimensions $k \times 1$ and $l \times 1$, respectively, and the null hypothesis to be tested is
$$
H_0: \beta_i(r) = 0, \quad\forall\, r = 1, \ldots, d, \quad\forall\, i = 1, \ldots, k,
$$
where $\beta_i(r)$ are the elements of the $\Bbeta(r)$. For a continuous factor of interest $k=1$ and $\Bbeta(r)$ serves as the test statistic. For a discrete factor of interest, in the default setup, $k$ is equal to the number of groups of the discrete factor, adding the additional condition that $\sum_i \beta_{i}(r)=0$ for all $r\in\{1, \ldots d\}$. Similarly, for interaction of a continuous and a discrete factor, $k$ is also equal to the number of groups of the categorical factor, adding the same additional condition. For the interaction of two discrete factors, $k$ is equal to the product of the numbers of groups of the discrete factors, adding the same additional condition.
If the argument \code{contrasts} of the function \fct{graph.flm} is set to \code{FALSE}, the test statistic for a discrete factor consist of $\beta_i(r)$ for all $r=1,\ldots,d$ and $i=1,\ldots,k$. This test allows to detect which groups deviate from the zero (mean).
An alternative test statistic is obtained by setting \code{contrasts = TRUE}: then the test statistic is formed by all the pairwise differences between the group effects, $\beta_i(r) - \beta_j(r)$ for all $r$ and $i\neq j$, $1\leq i < j \leq k$. This test specifies which specific groups are different in the post hoc nature. Note that this also holds for interaction terms.
Furthermore, all the options allow one to identify which of the components of the vector, $r\in\{1,\ldots,d\}$, lead to the potential rejection of the null hypothesis. Permutations under the null hypothesis are obtained using the Freedman-Lane procedure \citep{FreedmanLane1983, MrkvickaEtal2022, MrkvickaEtal2021a}.

Often factors are given for the whole function, i.e., they do not depend on argument $r$, and so the matrices $\mathbf X(r)$ and $\mathbf Z(r)$ are identical for every $r$. These kind of constant factors (such as sex and age in the considered example) can be provided in the argument \code{factors} of the \fct{graph.flm} function. However, this simplification is not necessary and factors varying in space can be provided in the argument \code{curve_sets}, along with the data curves in a named list.

The functional GLM is performed by the function \fct{graph.flm}:
\begin{Schunk}
\begin{Sinput}
R> res <- graph.flm(nsim = 999, formula.full = Y ~ Group + Sex + Age,
+    formula.reduced = Y ~ Sex + Age,
+    curve_sets = list(Y = abide_9002_23[['curve_set']]),
+    factors = abide_9002_23[['factors']], contrasts = TRUE,
+    GET.args = list(type = 'area'))
\end{Sinput}
\end{Schunk}
Here the arguments \code{formula.full} and \code{formula.reduced} specify the full GLM and the GLM where the interesting factor has been dropped out, and the number of simulations is given in \code{nsim}. Further arguments to \fct{global\_envelope\_test} can be passed in \code{GET.args}, e.g., the type of the global envelope.

\begin{figure}[ht!]
\centering
\includegraphics{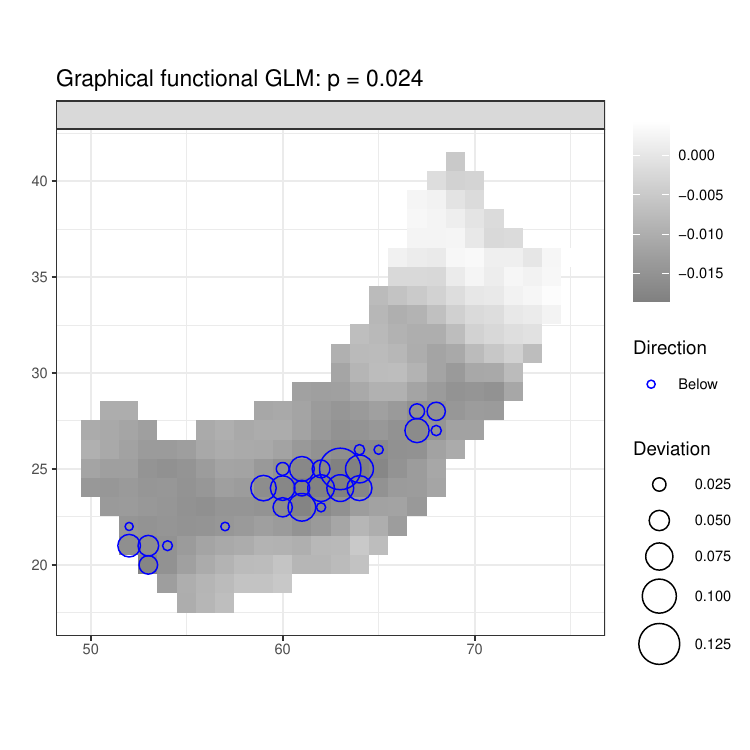}
  \caption{Graphical functional GLM for testing the effect of the group (autism, control) in the brain image example: the observed difference (autism-control) and the locations (blue circles) where the observed coefficient goes below the 95\% global envelope (area). The size of the circles are proportional to the size of the deviation of the observed function from the lower bound of the envelope divided by the width of the envelope.}
  \label{fig:flm_graph_abide}
\end{figure}

The \code{r} component of the \code{abide_9002_23[['curve_set']]} object is a data frame with columns \code{x}, \code{y}, \code{width} and \code{height}, where the \code{width} and \code{height} give the width and height of the pixels placed at \code{x} and \code{y}. When such two-dimensional argument values are provided in a \code{curve_set} object, the resulting default envelope plots produced by
\begin{Schunk}
\begin{Sinput}
R> plot(res) + scale_radius(range = 1.5 * c(1, 6))
\end{Sinput}
\end{Schunk}
are two-dimensional as well (Figure~\ref{fig:flm_graph_abide}). Above \fct{scale\_radius} function of \pkg{ggplot2} is used to enlarge the circles which show the size of deviation of the observed function from the bound (here lower) of the envelope divided by the width of the envelope.
Here only two groups were compared, and the plot shows that the brain measurements were lower in the autism group than in the control group in a part of the small example region (blue circles in Figure~\ref{fig:flm_graph_abide}).

When the basic assumption of the homoscedasticity in the linear model \eqref{eq:GLM} for every argument $r$ is violated, it is important to handle it. One possibility is to apply transformations to the functions a priori as suggested by \citet{MrkvickaEtal2020} and \citet{MrkvickaEtal2021a} (see Equation~\ref{eq:scaling}). Alternatively weighted least squares might be used for estimation of regression coefficients.

\subsubsection{Functional GLM based on $F$-statistics}

In the $F$-rank functional GLM, the same linear model (Equation~\ref{eq:GLM}) is fitted at each $r\in\{1,\ldots,d\}$ and permutations under the null hypothesis are obtained similarly by the Freedman-Lane procedure as in the graphical functional GLM \citep{FreedmanLane1983, MrkvickaEtal2022}.
However, the test statistic is the classical $F$~statistic \citep[see, e.g.,][]{WinklerEtal2014} which is calculated for the hypothesis that the data follows the simpler reduced model of the two proposed linear models that are nested within each other (given in \code{formula.full} and \code{formula.reduced}).
The use of the function \fct{frank.flm} is similar to that of \fct{graph.flm}:
\begin{Schunk}
\begin{Sinput}
R> res.F <- frank.flm(nsim = 999, formula.full = Y ~ Group + Age + Sex,
+    formula.reduced = Y ~ Age + Sex,
+    curve_sets = list(Y = abide_9002_23[['curve_set']]),
+    factors = abide_9002_23[['factors']], GET.args = list(type = 'area'))
R> plot(res.F, what = c("obs", "hi", "hi.sign"), sign.type = "contour")
\end{Sinput}
\end{Schunk}
For illustration of different options, we plot here the observed function (\code{what = "obs"}), the upper envelope (\code{what = "hi"}) and the observed function with the significant region (\code{what = "hi.sign"}). The significant region is here shown by contour lines, as chosen in the argument \code{sign.type}. Contours do not indicate the size of deviation from the bounds of the global envelope, but contours or colored regions (\code{sign.type = "col"}) can be preferable if the observed function exceeds the envelope on large parts of the domain.

\begin{figure}[ht!]
\centering
\includegraphics{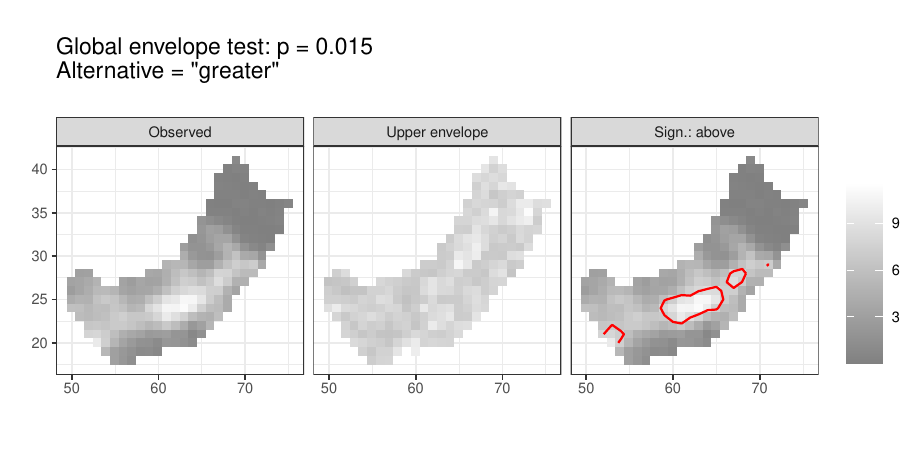}
  \caption{$F$-rank functional GLM for testing the effect of the group (autism, control) in the brain image example: the observed $F$~statistic, the upper bound of the one-sided 95\% global envelope (area), and the significant region (red contour lines) where the observed $F$~statistic exceeds the envelope.}
  \label{fig:flm_frank_abide}
\end{figure}

Figure~\ref{fig:flm_frank_abide} shows that the $F$-rank GLM found significant differences between the groups approximately at the same pixels $r\in\{1,\ldots,d\}$ of the brain image as the graphical functional GLM above.
In general, for a factor with more than two groups, the $F$-rank GLM is however not able to tell between which specific groups of the factor the differences occur (or which of the groups deviate from the mean).
In the case of heteroscedasticy, the weighted least squares test statistics can be used instead \citep{Christensen2002}.

\subsection{Confidence band in polynomial regression}\label{sec:confband}

\begin{figure}[ht!]
\centering
\includegraphics{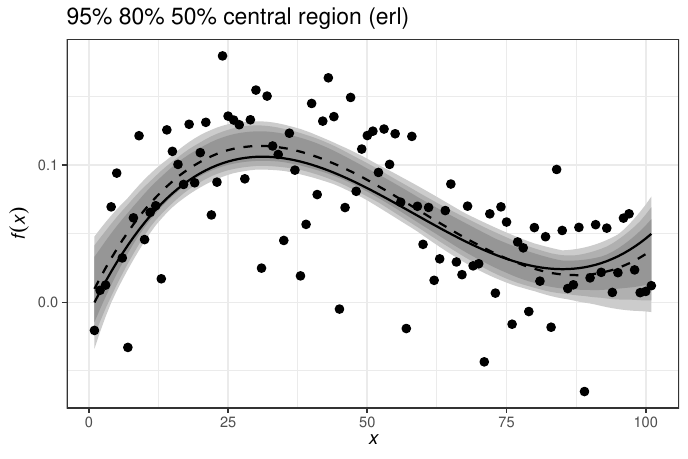}
  \caption{The global 95\%, 80\% and 50\% confidence bands (\code{'erl'}) for the cubic regression (nested gray bands; widest is 95\%), the true function (solid line) from which the data points (dots) were simulated and the median computed from the simulations (dashed line).}
  \label{fig:polynomial}
\end{figure}

The bootstrap procedure described in \citet{NarisettyNair2016} can be used to compute global confidence bands for the fitted curve in the linear or polynomial regression. In this example, regression data was simulated according to the cubic model $f(x) = 0.8x - 1.8x^2 + 1.05x^3$ for $x\in [0,1]$ with i.i.d.\ random noise (dots in Figure~\ref{fig:polynomial}). Then the data was fitted with cubic regression (black solid line in Figure~\ref{fig:polynomial}) and by permuting the residuals 2000 bootstrap samples were obtained and functions fitted \citep[see more details about the bootstrap procedure in][]{NarisettyNair2016}. Finally a \class{curve\_set} object was constructed of these bootstrapped functions and the \fct{central\_region} function was applied to this set to obtain the 95\%, 80\% and 50\% global confidence bands.
The multiple global bands were obtained by setting the argument \code{coverage = c(0.95, 0.80, 0.50)}.

The result of the procedure is shown in Figure~\ref{fig:polynomial}. The code can be found in the help file of the function \fct{central\_region} in \proglang{R}.

A theoretical 95\% confidence band can be considered under the given bootstrap scheme. Based on a simulation experiment where the theoretical confidence band was computed from 200000 bootstrapped functions, we observed that the 95\% confidence region computed as a convex hull from $s$ functions converged to the theoretical one from inside for increasing $s$. The 95\% confidence band computed as the extreme rank envelope from $s$ functions (see Equation~\ref{eq:kth_envelopes}) converged to the theoretical one from outside instead. Both these envelopes are finite approximations of the theoretical envelope. On the other hand, in the sense of Barnard's Monte Carlo test \citep{Barnard1963}, the global envelope test (convex hull) is exact for the given set of simulated functions. In the same sense, the confidence band reaches the given global level exactly under the given set of functions.


\section{Summary and discussion}\label{sec:summary}

We presented the \pkg{GET} package which was designed for global envelopes that are constructed for a general vector and have IGI (Definition~\ref{IGI}). The great advantage of these methods is their graphical output given by the $100(1-\alpha)$\% (e.g., 95\%) central region, which helps one to interpret the results in the various applications. The package implements different types of global envelopes (see Table~\ref{tab:GEoverview}) and their usage in general and for several specific problems.
Because the global envelopes can be used for so many different purposes specified in cases (i)-(iii) in Section~\ref{sec:intro}, there are several other software, particularly other \proglang{R} packages, that deal with methods that can be used for the same purposes as the methods in \pkg{GET}. However, to the best of our knowledge, \pkg{GET} is the first package specializing to the global envelopes with IGI.
Besides, the \pkg{GET} package can provide testing results both under the FWER control and the FDR control.

Besides the graphical interpretation, another advantage of the proposed global rank envelopes is their non-parametric (rank-based) nature, which ensures that the functional or multivariate data coming into the analysis can be inhomogeneous across the domain of their arguments and this phenomenon does not influence the result of the analysis. For example, before the methods discussed in this paper appeared, formal goodness-of-fit testing in spatial statistic was commonly based on the unscaled MAD test \citep{Ripley1981} or its non-graphical integrated counterpart \citep{Diggle1979}. However, the result of these tests is influenced by unequal variability of the test function across its domain leading in general to loss of power \citep{MyllymakiEtal2015, MyllymakiEtal2017}.
A similar situation appears in the permutation GLM tests which, in the functional data analysis or neuroimage analysis \citep[see, e.g.,][]{WinklerEtal2014}, are commonly based on the $F$~statistic that standardizes the first and second moments of the data but not the high quantiles. Thus, when the data are inhomogeneous across the domain and non-normal, the commonly used $F$~max test (which is similar to the unscaled MAD test) is influenced by the inhomogeneous quantiles. The rank-based tests discussed here can then lead to higher power \citep[for details see][]{MrkvickaEtal2022}. Similarly, the rank based methods can adjust the shape of the central region to inhomogeneous distribution of the studied functions. Therefore, the global rank envelopes are a valuable tool in all these situations.

A further advantage of the rank tests is that it allows one to give equal weights to the components fed in. Thus, the method is particularly well suited also for multiple testing with several univariate or functional test statistics \citep{MrkvickaEtal2017}, for constructing central regions jointly for various transformations of the functions \citep[for details see][]{DaiEtal2020}, and for combining various dimensions of multidimensional functions or various functional elements of multivariate functions. Further, it can be used for functional cluster analysis so that various transformations of the functions have the same impact on the clustering algorithm \citep{DaiEtal2022}.

Finally, the good properties of the methods presented here are retained also in the case of testing a composite hypothesis: the two-stage Monte Carlo test is applicable to these graphical methods (Appendix \ref{app:adjustedGE}).

We are committed to developing the \pkg{GET} package further. For example, new types of global envelopes can be added, if such are invented, and support for specific applications or different type of data will be extended.

\section*{Acknowledgments}

MM was financially supported by the Research Council of Finland (project numbers 295100, 306875, 327211) and TM by the Grant Agency of Czech Republic (Project No.\ 19-04412S and 24-11096S).
The authors thank Ji\v{r}\'{i} Dvo\v{r}\'{a}k, Pavel Grabarnik, Ute Hahn, Mikko Kuronen, Michael Rost and Henri Seijo for their contributions and suggestions of code, as well as Stavros Athanasiadis, Wenlin Dai, Marc G.\ Genton, Milan J\'ilek, Naveen Naidu Narisetty, Tom\'a\v s Roskovec, Samuel Soubeyrand and Ying Sun for their collaboration on developing global envelope methods.
The authors wish to acknowledge CSC -- IT Center for Science, Finland, for computational resources.




\appendix

\section{Definitions of global envelopes}\label{app:GEdef}

Here, we provide definitions of the different global envelopes and the corresponding measures, as outlined in Table \ref{tab:GEoverview}.
When working with functions, it is assumed that they have already been discretized as required in practice.
Thus, we consider general vectors $\TT_i=(T_{i1},\ldots,T_{id})$, $i=1,\ldots,s$, for which a global envelope is to be constructed.

The measures satisfy the IGI property with probability 1 if there are no pointwise ties, meaning that there are no ties in $T_{1k},\ldots,T_{sk}$ for every $k = 1,\ldots,d$.
This holds if $\TT_i$ are realizations of an absolutely continuous random vector.
If pointwise ties do occur, the IGI property may be violated in elements of the vector where ties are present.
For the measures based on pointwise ranks (Sections \ref{sec:rank}-\ref{sec:area}), we use the pointwise mid-rank for tied values to weigh down the influence of elements ($k$) where many $T_{ik}$, $i=1,\ldots,s$, coincide.

\subsection{Global rank envelope}\label{sec:rank}

The extreme rank $R_i$ of the vector $\TT_i$ is defined as the minimum of its pointwise ranks, namely
  \begin{equation}\label{eq:minglobalrank}
  R_i=\min_{k=1,\ldots,d} R_{ik},
  \end{equation}
where the pointwise rank $R_{ik}$ is the rank of the element $T_{ik}$ among the corresponding elements $T_{1k}, T_{2k}, \ldots, T_{s k}$ of the $s$ vectors such that the lowest ranks correspond to the most extreme values of the statistics.
How the pointwise ranks are determined depends on whether a one-sided or a two-sided global envelope (test) is to be constructed: Let $r_{1k}, r_{2k}, \ldots, r_{s k}$ be the raw ranks of $T_{1k}, T_{2k}, \ldots, T_{s k}$, such that the smallest $T_{ik}$ has rank 1. In the case of ties among $T_{1k}, T_{2k}, \ldots, T_{s k}$, the raw ranks are averaged. The pointwise ranks are then calculated as
\begin{equation*}
  R_{ik}=\begin{cases}
    r_{ik}, &\text{for the one-sided case 'less',} \\
    s+1-r_{ik}, &\text{for the one-sided case 'greater',} \\
   \min(r_{ik}, s+1-r_{ik}), &\text{for the two-sided case}.
  \end{cases}
\end{equation*}
The cases 'less' and 'greater' are most often relevant in a hypothesis testing case (ii) (see Section~\ref{sec:intro}).
The case 'less' corresponds to the alternative that the values of interesting vector ($\TT_1$) are smaller than under the null hypothesis. 
The case 'greater' corresponds to the alternative that the values are larger than under the null hypothesis.
The extreme rank measure $R_i$ induces an ordering of $\TT_i$, $i=1,\ldots,s$, which can be used to detect the extremeness of the vectors among each other. Given that $\TT_1$ is the observed vector in the Monte Carlo or permutation test, the (conservative) $p$~value of the test is equal to $p_+ = \sum_{i=1}^{s} \1 (R_i \leq R_1)  \big/ s$.
Since the extreme ranks can have many ties, the test is also equipped with the liberal $p$~value, $p_- =  \sum_{i=1}^{s} \1 (R_i < R_1) \big/ s$. Then, when $\alpha$ falls inside the $p$~interval $(p_-,p_+]$, the decision of the test is not defined.

The $100(1-\alpha)$\% global rank envelope induced by this measure is defined through
\begin{equation}\label{eq:kth_envelopes}
\TT^{\text{local}, l}_{\lo \, k}= \underset{i=1,\ldots,s}{{\min}^{l}} T_{i k}
\quad\text{and}\quad
\TT^{\text{local}, l}_{\up \, k}= \underset{i=1,\ldots,s}{{\max}^{l}} T_{i k} \quad \text{for } k = 1, \ldots , d,
\end{equation}
by taking $l=R_{(\alpha)}$, according to the point 1.\ of IGI (see Definition~\ref{IGI}), i.e., setting $\TT^{(\alpha)}_{\lo \, k} = \TT^{\text{local}, R_{(\alpha)}}_{\lo \, k}$ and $\TT^{(\alpha)}_{\up \, k} = \TT^{\text{local}, R_{(\alpha)}}_{\up \, k}$. Here $\min^l$ and $\max^l$ denote the $l$-th smallest and largest values, respectively, and $l=1,2,\ldots,\lfloor s/2\rfloor$.
If $\TT_i$ is strictly outside the envelope for some $k=1,\ldots,d$, then also $R_i < R_{(\alpha)}$, and if $\TT^{(\alpha)}_{\lo \, k}<\TT_{ik}<\TT^{(\alpha)}_{\up \, k}$ for all $k=1,\ldots,d$, then $R_i > R_{(\alpha)}$. However, if $\TT_i$ coincides either with $\TT^{(\alpha)}_{\lo \, k}$ or $\TT^{(\alpha)}_{\up \, k}$ for some $k=1,\ldots,d$, then $R_i = R_{(\alpha)}$, and $\alpha \in (p_-,p_+]$ in the testing case.

Because the extreme rank can achieve many ties \citep[see, e.g.,][]{MrkvickaEtal2022}, it is necessary to use a relatively large $s$ for the global rank envelope.
The following three refinements of the extreme rank solve the ties problem and enable the use of a smaller $s$.

\subsection{Global extreme rank length (ERL) envelope}\label{sec:erl}

The extreme rank length (ERL) measure \citep{MyllymakiEtal2017, NarisettyNair2016} refines the extreme rank measure by breaking the ties in the extreme ranks $R_i$ by taking into account also the number of $R_{ik}$ which are equal to $R_i$. Further, the numbers of ranks equal to $R_i+1$, $R_i+2$, ... are used to break any remaining ties. 

Formally, the ERL measure of $\TT_i$ is defined based on the vector of the pointwise ordered ranks
$\mathbf{R}_i=(R_{i[1]}, R_{i[2]}, \ldots , R_{i[d]})$, where the ranks are arranged from smallest to largest, i.e., $R_{i[k]} \leq R_{i[k^\prime]}$ whenever $k \leq k^\prime$. While the extreme rank (Equation~\ref{eq:minglobalrank}) corresponds to $R_i=R_{i[1]}$, the ERL measure takes all these ranks into account by the reverse lexical ordering. 
The ERL measure of $\TT_i$ is
\begin{equation}
\label{eq:lexicalrank}
   E_i = \frac{1}{s}\sum_{i^\prime=1}^{s} \1(\mathbf{R}_{i^\prime} \prec \mathbf{R}_i)
\end{equation}
where
\[
\mathbf{R}_{i^\prime} \prec \mathbf{R}_{i} \quad \Longleftrightarrow\quad
  \exists n\leq d: R_{i^\prime[k]} = R_{i[k]} \forall k < n, R_{i^\prime[n]} < R_{i[n]}.
\]
The division by $s$ leads to normalized ranks that obtain values between 0 and 1. Consequently, the ERL measure corresponds to the extremal depth of \citet{NarisettyNair2016}.

The probability of having a tie in the ERL measure is rather small, thus practically the ERL solves the ties problem. The final $p$~value of a Monte Carlo test is $p_\text{erl}=  \sum_{i=1}^{s} \1 (E_i \leq E_1)  \big/ s$.

Let $E_{(\alpha)}$ be defined according to the point 1.\ of IGI and  $I_\alpha = \{i\in 1,\ldots, s: E_i \geq E_{(\alpha)} \}$ be the index set of vectors less or as extreme as $E_{(\alpha)}$.
Then the $100(1-\alpha)$\% global ERL envelope induced by $E_i$ is
\begin{equation}\label{erl_envelope}
\TT^{(\alpha)}_{\lo \, k}= \underset{i \in I_\alpha}{{\min}}\ T_{i k}
\quad\text{and}\quad
\TT^{(\alpha)}_{\up \, k}= \underset{i \in I_\alpha}{{\max}}\ T_{i k} \quad \text{for } k = 1, \ldots , d,
\end{equation}
see \citet{NarisettyNair2016} and \citet{MrkvickaEtal2020}.

\subsection{Global continuous rank envelope}\label{sec:cont}

The ties can alternatively be broken by the continuous rank measure \citep{Hahn2015, MrkvickaEtal2022} which refines the extreme rank measure by considering instead of the (discrete) pointwise ranks $R_{ik}$ continuous pointwise ranks $C_{ik}$ defined by the ratios of $T_{ik}$ to the closest other $T_{jk}$, $j=1,\ldots,s$, $j\neq i$.

Formally, the continuous rank measure is
$$C_i = \frac{1}{s}\min_{k=1,\ldots, d} C_{ik},$$
where $s$ scales the values to interval from 0 to 1.
The definition of pointwise continuous ranks $C_{ik}$ depends again on whether a one-sided or two-sided global envelope (test) is to be constructed:
\begin{equation*}
  C_{ik}=\begin{cases}
    c_{ik}, &\text{for the one-sided case 'less'} \\
    s-c_{ik}, &\text{for the one-sided case 'greater'} \\
   \min(c_{ik}, s-c_{ik}), &\text{for the two-sided case}.
  \end{cases}
\end{equation*}
where $c_{ik}$ is the raw continuous rank of $T_{ik}$ among $T_{1k}, \ldots, T_{sk}$ according to Definition~\ref{contrank} and the three cases are similar to those of $R_{ik}$ above.

\begin{definition}\label{contrank}
Let $y_{[1]} \le y_{[2]} \le \ldots \le y_{[s]}$ denote the ordered set of values $y_i, i=1,2, \ldots, s$.
Define the raw continuous rank such that the smallest $y_i$ has smallest rank following \citet{MrkvickaEtal2022}:
\begin{equation*}
c_{[j]} = j - 1 + \frac{y_{[j]}-y_{[j-1]}}{y_{[j+1]}-y_{[j-1]}}
\end{equation*}
for $j = 2, 3, \ldots, s-1$ and
\begin{align*}
c_{[1]} = \exp\left(-\frac{y_{[2]}-y_{[1]}}{y_{[s]}-y_{[2]}}\right) , \quad
c_{[s]} =  s - \exp\left(-\frac{y_{[s]}-y_{[s-1]}}{y_{[s-1]}-y_{[1]}}\right).
\end{align*}
If there are ties, $y_{[i-1]} < y_{[i]} = \ldots = y_{[j]} < y_{[j+1]}$, then the raw continuous rank is defined as $c_{[k]} = \frac{i+j}{2} - \frac{1}{2}$ for $k = i, i+1, \ldots, j$.
\end{definition}

The $p$~value of the univariate Monte Carlo test is $p_\text{cont}= \sum_{i=1}^{s} \1 (C_i \leq C_1)  \big/ s$.
The $100(1-\alpha)$\% global continuous rank envelope induced by $C_i$ is constructed in the same way as global ERL envelope, i.e., as a hull of $\TT_i$ for which $C_i\geq C_{(\alpha)}$, where $C_{(\alpha)}$ is defined according to the point 1.\ of IGI. This is achieved by having $I_\alpha = \{i\in 1,\ldots, s: C_i \geq C_{(\alpha)} \}$ in Equation~\ref{erl_envelope}.

\subsection{Global area rank envelope}\label{sec:area}

Another refinement of rank envelope is the area rank measure \citep{MrkvickaEtal2022},
%
$$A_i = \frac{1}{s}\left(R_i - \frac{1}{d}\sum_{k=1}^d (R_i-C_{ik}) \1(C_{ik} < R_i)\right).$$
The area measure breaks the ties in the extreme ranks by the sum (area) of the differences between the extreme rank $R_i$ and the pointwise continuous rank $C_{ik}$ from those $k=1,\ldots,d$ where the continuous rank is smaller than the extreme rank.
The univariate Monte Carlo test is performed based on $A_i$ with $p_\text{area}= \sum_{i=1}^{s} \1 (A_i \leq A_1)  \big/ s$.
The $100(1-\alpha)$\% global area rank envelope induced by $A_i$ is constructed similarly as the global ERL and continuous rank envelopes, with $I_\alpha = \{i\in 1,\ldots, s: A_i \geq A_{(\alpha)} \}$ in Equation~\ref{erl_envelope}.

\subsection{Global directional quantile, studentized and unscaled envelope}

The above four global envelopes are based on the whole distributions of $T_{1 k}, \ldots, T_{s k}$, $k=1, \ldots , d$. It is also possible to approximate the distribution by a few sample characteristics. The sample characteristics are in the package \pkg{GET} estimated from $T_{i k}$, $i=1,\ldots,s$, for each $k$.

{\em The global directional quantile envelope} uses the expectation $T_{0k}$, $\beta\%$ upper quantile $\overline{T}_{\cdot k}$ and $\beta\%$ lower quantile $\underline{T}_{\cdot k}$ to approximate the distributions. Setting $\beta=2.5$ was used in \citet{MyllymakiEtal2017}; setting $\beta=25$ can also be useful especially for defining the 50\% central region from a low number of functions. Note that $\beta$ has to be greater than $100/s$ in order to be able to estimate the $\beta$ and $1-\beta$ quantiles. The directional quantile measure \citep{MyllymakiEtal2015, MyllymakiEtal2017} $D_i$ is defined as
\begin{equation}\label{Dinfty_qdir}
D_i = \max_{k} \left( \1(T_{ik} \geq T_{0k}) \frac{ T_{ik} - T_{0k} }{ |\overline{T}_{\cdot k} - T_{0k}| } + \1 (T_{ik} < T_{0k}) \frac{ T_{0k} - T_{ik} }{ |\underline{T}_{\cdot k}-T_{0k}| } \right),
\end{equation}
From historical reasons, $D_i$ is defined to be bigger for more extreme vectors. The same holds for the following two measures.
The univariate Monte Carlo test is performed based on $D_i$ with $p_\text{qdir}= \sum_{i=1}^{s} \1 (D_i \geq D_1)  \big/ s$, and the $100(1-\alpha)$\% global directional quantile envelope induced by $D_i$ is defined by
\begin{equation}
\TT^{(\alpha)}_{\lo \, k}= T_{0k} - D_{(\alpha)} |\underline{T}_{\cdot k}-T_{0k}|
\quad\text{and}\quad
\TT^{(\alpha)}_{\up \, k}= T_{0k} + D_{(\alpha)} |\overline{T}_{\cdot k}-T_{0k}| \quad \text{for } k = 1, \ldots , d,
\end{equation}
where $D_{(\alpha)}$ is taken according to the point 1.\ of IGI.

{\em The global studentized envelope}
approximates the distribution instead by the expectation $T_{0k}$ and the standard deviation $\text{sd}(T_{\cdot k})$.
The studentized measure \citep{MyllymakiEtal2015, MyllymakiEtal2017} is
\begin{equation}\label{Dinfty_st}
S_i = \max_{k} \Big| \frac{ T_{ik} - T_{0k}}{ \text{sd}(T_{\cdot k}) } \Big| ,
\end{equation}
and the univariate Monte Carlo test is performed based on $s_i$ with $p_\text{st}= \sum_{i=1}^{s} \1 (S_i \geq S_1) \big/ s$.
The $100(1-\alpha)$\% global studentized envelope induced by $S_i$ is defined by
\begin{equation}
\TT^{(\alpha)}_{\lo \, k}= T_{0k} - S_{(\alpha)} \text{sd}(T_{\cdot k})
\quad\text{and}\quad
\TT^{(\alpha)}_{\up \, k}= T_{0k} + S_{(\alpha)} \text{sd}(T_{\cdot k})  \quad \text{for } k = 1, \ldots , d,
\end{equation}
where $S_{(\alpha)}$ is taken according to the point 1.\ of IGI.

{\em The global unscaled envelope}
considered for the sake of completeness has its origin in the classical Kolmogorov-Smirnov statistic. 
The unscaled measure $U_i$ can be defined as $U_i = \max_{k} |T_{ik} - T_{0k}|$, the univariate Monte Carlo test performed based on $U_i$ has the $p$~value $p_\text{unsc}= \sum_{i=1}^{s} \1 (U_i \geq U_1)  \big/ s$, and the $100(1-\alpha)$\% global unscaled envelope induced by $u_i$ is given by
\begin{equation}
\TT^{(\alpha)}_{\lo \, k}= T_{0k} - U_{(\alpha)}
\quad\text{and}\quad
\TT^{(\alpha)}_{\up \, k}= T_{0k} + U_{(\alpha)} \quad \text{for } k = 1, \ldots , d,
\end{equation}
where $U_{(\alpha)}$ is taken according to the point 1.\ of IGI.
A problem of this envelope is that its width is the same along the whole domain, thus it cannot account for the changes in the variability of the distributions $\TT_i$ across different values of $k$ \citep{MyllymakiEtal2015, MyllymakiEtal2017}. 

\section{Combined global envelopes}\label{app:combinedGE}

Assume that there are $G$ vectors $\TT_i^j = (T_{i1}^j, \ldots , T_{i{d_j}}^j), j=1, \ldots , G$, $i=1, \ldots , s$, $d_j\geq 1$, and the construction of a global envelope is wanted jointly for all of them. A combined global envelope test can be made in two different ways.

In the \emph{two-step combining procedure}, 
first, a measure is chosen for each $j=1, \ldots, G$ and computed for the vectors $\TT_i^j$, $i=1,\ldots,s$ and $j=1,\ldots, G$. Let the resulting measures be $m_i^j$. 
As the second step, the one-sided extreme rank length is applied to the new vector $\TT_i' = (m_i^1, m_i^2, \ldots, m_i^G)$ of the measures.
As a result, a joint sorting of vectors $\TT_i^1, \ldots \TT_i^G, i=1, \ldots, s$, is obtained and a joint extreme rank length measure $E_i$ is attached to every $i=1, \ldots , s$.  
The $p$~value of the combined Monte Carlo test is $p_\text{erl}=  \sum_{i=1}^{s} \1 (E_i \leq E_1)  \big/ s$, and the common $100(1-\alpha)$\% global envelope is constructed similarly as the $100(1-\alpha)$\% global extreme rank length envelope (Equation~\ref{erl_envelope}): Let $E_{(\alpha)}$ be defined according to the point 1.\ of IGI and $I_\alpha = \{i\in 1,\ldots, s: E_i \geq E_{(\alpha)} \}$ be the index set of vectors less or as extreme as $E_{(\alpha)}$.
Then the common $100(1-\alpha)$\% global envelope is
\begin{equation}
\TT^{(\alpha),j}_{\lo \, k}= \underset{i \in I_\alpha}{{\min}}\ T_{i k}^j
\quad\text{and}\quad
\TT^{(\alpha), j}_{\up \, k}= \underset{i \in I_\alpha}{{\max}}\ T_{i k}^j \quad \text{for } k = 1, \ldots , d_j, j = 1, \ldots , G.
\end{equation}

The extreme rank length measure is chosen in the second step because it gives the same weight to every component (even when $d_j, j=1, \ldots , G$, are different or even if different measures are used in the first step), it is based on ranks only and it achieves almost no ties.

In cases where $d_1 = \ldots = d_G$ ($= d$), it is also possible to use a simple {\it one-step combining procedure}. 
Then the global envelope (any of those in Table~\ref{tab:GEoverview}) is constructed for the long vectors 
$$\TT_i = (T_{i1}^1, \ldots , T_{id}^1, T_{i1}^2, \ldots  , T_{id}^2, \ldots \ldots , T_{i1}^G, \ldots , T_{id}^G),\quad i=1, \ldots , s.$$
The one-step combining can be used for example when multivariate functional data $(\mathcal T_{i1}, \ldots , \mathcal T_{id})$, $i=1, \ldots , s$, where $\mathcal T_{ik}=(T_{ik}^1, \ldots , T_{ik}^{J})$ are multivariate vectors of $J$ elements, are investigated. 
Then it is possible to separate the dimensions into a set of $J$ marginal vectors applying the one-step combining procedure, i.e., to take $\TT_i = (T_{i1}^1,\ldots , T_{id}^{1}, \ldots , T_{i1}^J, \ldots , T_{id}^{J})$.
Further, it is also possible to add other vectors expressing the correlation between the elements of the vectors,\ e.g., if we have a two-dimensional functional data, the vector $(T_{i1}^1 T_{i1}^2 - T_{01}^1T_{01}^2, \ldots , T_{id}^1 T_{id}^2-T_{0d}^1T_{0d}^2)$ can be added into $\TT_i$ behind the marginal vectors. Here $T_{0k}^j$ denotes the expectation of $T_{i k}^j$, $i=1,\ldots,s$.

The graphical functional ANOVA and GLM (see the functions in Table~\ref{tab:FUNCTIONSoverview}) use the one-step combining procedure to merge the mean or contrast vectors under inspection, because in this case all the vectors have the same structure \citep[see Sections~\ref{sec:fANOVA} and \ref{sec:fGLM} and][]{MrkvickaEtal2020, MrkvickaEtal2021a}. On the other hand, for generality, the default combining procedure of global envelope construction functions in \pkg{GET} is the two-step procedure, which is presented for the first time here as an improvement of the combined tests of \citet{MrkvickaEtal2017} (see an example in Section~\ref{sec:cr}). The combined envelopes are implemented in the \fct{central\_region} and \fct{global\_envelope\_test} functions as mentioned above, and the one- or two-step procedure can be specified in the argument \code{nstep} (either 1 or 2).

\section{Adjusted global envelopes for composite null hypotheses}\label{app:adjustedGE}

The Monte Carlo tests for which the global envelopes are constructed are exact only in the case when the null hypothesis is simple, i.e., when no parameters have to be estimated. This is the case in permutation tests of task (iii), but in task (ii) the null hypothesis can often be composite, i.e., some parameters of the null model have to be estimated. In such a composite case, the classical Monte Carlo test can be liberal or conservative. This problem can be solved by a two-stage procedure, where in the first stage the level of the test is estimated. Such a procedure was first introduced by \citet{DaoGenton2014} for Monte Carlo tests. \citet{MyllymakiEtal2017} extended this adjusted method for global envelopes. \citet{BaddeleyEtal2017} improved the procedure further in order to obtain an exact significance level. Here the procedure of \citet{BaddeleyEtal2017} is summarized and extended for global envelopes as implemented in \pkg{GET}.

Let $M$ denote the chosen measure and $\alpha$ the chosen significance level. Let $\TT_1$ be the test vector computed from the data.
\begin{enumerate}
\item Estimate the parameters $\theta_1$ of the null model.
\item Simulate $s_2-1$ replicates of the data from the null model with the estimated parameters $\hat{\theta}_1$, and compute the test vectors $\TT_1^1 = \TT_1, \TT_2^1, \ldots , \TT_{s_2}^1$ (create a \class{curve\_set} object of vectors, $C_1$).
\item Simulate another $s-1$ replicates of the data from the null model with the parameters $\hat\theta_1$ and estimate the parameters of the null model from each of them ($\hat{\theta}_i, i=2, \ldots , s$),
\item For every $i=2, \ldots, s$, simulate $s_2-1$ replicates from the null model with parameters $\hat{\theta}_i$, and compute the test vectors $\TT_1^i, \TT_2^i, \ldots , \TT_{s_2}^i$ (create a \class{curve\_set} object of vectors, $C_i$).
\item For each set of curves $C_i$, $i=1,2,\ldots,s$, compute the Monte Carlo $p$~value 
$p_i= \sum_{j=1}^{s_2} \1 (M_j^i \leq M_1^i)  \big/ s_2$,
where $M_1^i, \ldots , M_{s_2}^i$ are the chosen measure computed for $\TT_1^i, \ldots , \TT_{s_2}^i$.
\item The adjusted MC $p$~value is $p_\text{adj}= \sum_{j=1}^{s} \1 (p_i \leq p_1)  \big/ s$.
\item Let $p^\alpha$ denote the lower $\alpha$ quantile of the sample $p_1, \ldots , p_s$. Construct the chosen $100(1-p^\alpha)$\% global envelope from $\TT_1^1, \ldots , \TT_{s_2}^1$.
\end{enumerate}
This adjusted test is implemented in the \fct{GET.composite} function of the \pkg{GET} package.
Examples can be found in the help page of the function in \proglang{R} and in Section~\ref{sec:GOTcomposite}.

\end{document}